%% file: main.tex
\documentclass[10pt,twocolumn]{z-article} %
\input{z-package}
\pdfoutput=1
\begin{document}

\input{macros}
\input{title}

\input{abstract}
\input{intro}

\input{bg}

\input{ovr} 
\input{m-char} 
\input{m-brk} 
\input{m-pred} 
\input{m-inter} 
\input{m-tier} 
\input{dis} 
\input{rel}
\input{conclude}

\input{ack}

\input{main.bbl}

\end{document}

%% file: z-package.tex
\usepackage{S/oneinchmargins}  
\usepackage{times}
\usepackage{relsize}
\usepackage{enumerate}
\usepackage{enumitem}
\usepackage{calc}
\usepackage{graphicx}
\usepackage[square,sort&compress,comma,numbers]{natbib}
\usepackage{url}  
\usepackage{xurl}
\usepackage[dvipsnames]{xcolor}
\usepackage{hyperref}

\hypersetup{colorlinks=true,
citecolor=Maroon,
linkcolor=Green,
urlcolor=Maroon}

\usepackage{fixltx2e}
\usepackage{amsmath}

\usepackage{setspace}
\usepackage{rotating}
\usepackage{xspace}

\usepackage{floatflt}
\usepackage{wrapfig}
\usepackage{alltt}
\usepackage{epstopdf}
\usepackage{subcaption}

\usepackage{fancyvrb}

\usepackage{etoolbox}

\usepackage{colortbl} 
\usepackage{tcolorbox}

\usepackage{wasysym}  
\usepackage{pifont}   

\usepackage{upgreek}

\usepackage{tikz, pgfplots, pgfplotstable}
\usetikzlibrary{patterns}
\usepgfplotslibrary{fillbetween}
\usepackage{upgreek}
\usepackage{bm}

%% file: macros.tex
\def \numWorkloads {265\xspace} 
\def \numLatLevels {7\xspace} 
\def \numCxlDevs {4\xspace} 
\def \numCpus {4\xspace} 
\def \cxlLatMin {140\xspace} 
\def \cxlLatMax {410\xspace} 
\def \numPmuCounters {12\xspace} 
\def \pctTierImprv {177\%\xspace} 
\def \pctInterleaveImprv {13\%\xspace} 

\def \modelAccEmrNuma {99\%\xspace} %
\def \modelAccEmrCXLA {94\%\xspace} %
\def \modelAccEmrCXLB {91\%\xspace} %
\def \modelAccSKXNuma {97\%\xspace} %
\def \modelTgt {10\%\xspace} %

\def \cxla {CXL-A\xspace}
\def \cxlb {CXL-B\xspace}
\def \cxlc {CXL-C\xspace}
\def \cxld {CXL-D\xspace}

\def \alto {Alto\xspace} 
\def \znuma {zNUMA\xspace} 

\def \pone {$P_{1}$}
\def \ptwo {$P_{2}$}
\def \pthr {$P_{3}$}
\def \pfou {$P_{4}$}
\def \pfiv {$P_{5}$}
\def \psix {$P_{6}$}
\def \psev {$P_{7}$}
\def \peig {$P_{8}$}
\def \pnin {$P_{9}$}
\def \pten {$P_{10}$}
\def \pele {$P_{11}$}
\def \ptwet {$P_{12}$}
\def \pthit {$P_{13}$}
\def \pfout {$P_{14}$}
\def \pfift {$P_{15}$}
\def \psixt {$P_{16}$}

\def \ns {ns\xspace}
\def \bdr {}


\def \sys {SupMario\xspace}
\makeatother

\def \ione {I$_1$}

\def \us {$\upmu$s\xspace}  
\def \usb {\mathbf{\upmu}\text{s}}

\def \yes {$\surd$}  
\def \cm {\checkmark}

\def \none {N$_1$}
\def \ntwo {N$_1$}
\def \ntri {N$_1$}
\def \nnnn {N$_n$}

\def \ra {$\rightarrow$}

\def \tms {\texttimes\xspace}
\def \roughly {{\small $\sim$}}
\def \lmt {{\raisebox{.25ex}{\scriptsize {\textgreater}}}}
\def \mlt {{\raisebox{.25ex}{\scriptsize {\textless}}}}

\def \whitecircle {$\ocircle$}
\def \blackcircle {\ding{108}}
\def \whitesquare {$\Box$}
\def \blacksquare {\ding{110}}
\def \whitediamond {$\Diamond$}
\def \blackdiamond {\ding{117}}
\def \whitetriangle {$\bigtriangleup$}
\def \blacktriangle {\ding{115}}
\def \whitedtriangle {$\bigtriangledown$}
\def \blackdtriangle {\ding{116}}

\newcommand{\tc}[1]{\textbf{\textcircled{#1}}}
\def \tcone {\tc{1}\xspace}
\def \tctwo {\tc{2}\xspace}
\def \tctri {\tc{3}\xspace}
\def \tcfour {\tc{4}\xspace}
\def \tcfive {\tc{5}\xspace}

\newcommand{\vtwenty}{\vspace{20pt}}
\newcommand{\vfifteen}{\vspace{15pt}}
\newcommand{\vten}{\vspace{10pt}}
\newcommand{\vfive}{\vspace{5pt}}
\newcommand{\vthree}{\vspace{3pt}}
\newcommand{\vtwo}{\vspace{2pt}}

\newcommand{\vminfive}{\vspace{-5pt}}
\newcommand{\vminten}{\vspace{-10pt}}
\newcommand{\vminfifteen}{\vspace{-15pt}}
\newcommand{\vmintwenty}{\vspace{-20pt}}

\def \hmina {\hspace{-0.1in}}
\def \hminb {\hspace{-0.2in}}

\newcommand{\ub}[1]{\underline{{\bf #1}}}
\newcommand{\ts}[1]{{\tt{\small#1}}}
\newcommand{\tts}[1]{{\tt{\footnotesize#1}}}

\newcommand{\bquote}{\vspace{-0.25cm} \begin{quote}}
\newcommand{\equote}{\end{quote}\vspace{-0.2cm} }
\def \sec {\S}
\def \yes {$\surd$}

\def \nospace {
  \setlength{\itemsep}{0pt}
  \setlength{\parskip}{0pt}
  \setlength{\parsep}{0pt}
}

\newcommand{\myquote}[1]{
\begin{quote}
\centering
\small
\textit{#1}
\end{quote}
}

\newenvironment{enumerate2}{
\begin{enumerate}[leftmargin=*] \vminfive
  \setlength{\itemsep}{2pt}
  \setlength{\parskip}{0pt}
  \setlength{\parsep}{0pt}
}{
\end{enumerate}
\vminfive
}

\newenvironment{itemize2}{
    \begin{itemize}[leftmargin=8pt] \vminfive
        \setlength{\itemsep}{2pt}
        \setlength{\parskip}{0pt}
        \setlength{\parsep}{0pt}
}{
  \end{itemize}
  \vminfive
}


\newcommand{\notes}[1]{\textcolor{darkgray}{{\footnotesize {\em (Notes: #1)}}}}

\newcommand{\student}[1]{\textcolor{purple}{{\footnotesize {\bf (STU: #1)}}}}

\newcounter{hsgcounter}
\newcommand{\hsg}[1]{{\footnotesize
\textbf{\textcolor{red}{(HSG$_{\arabic{hsgcounter}}$: #1)}}}
\stepcounter{hsgcounter}}

\newcommand{\revs}[1]{\textcolor{blue}{\textit{(REVS: #1)}}}
\newcommand{\hcl}[1]{{\footnotesize\textbf{\textcolor{magenta}{        [ HCL: #1 ]}}}}
\newcommand{\jsl}[1]{{\footnotesize\textbf{\textcolor{CornflowerBlue}{ [ JSL: #1 ]}}}}
\newcommand{\hah}[1]{{\footnotesize\textbf{\textcolor{NavyBlue}{ [ HAH: #1 ] }}}}
\newcommand{\cwm}[1]{{\footnotesize\textbf{\textcolor{blue}{           [ CWM: #1 ] }}}}

\newcommand{\pc}[1]{\textcolor{blue}{\textit{(PC: #1)}}} 
\newcommand{\todo}[1]{\textcolor{red}{{\footnotesize {\bf (TODO: #1)}}}}

\newcommand{\newtxt}[1]{\textcolor{blue}{#1}} 
\newcommand{\oldtxt}[1]{\textcolor{gray}{{\footnotesize {\em OLD TEXT: #1}}}}
\newcommand{\bluetxt}[1]{\textcolor{blue}{#1}}
\newcommand{\rbt}[1]{\textcolor{red}{\textbf{#1}}}
\newcommand{\bbt}[1]{\textcolor{blue}{\textbf{#1}}}

\def \vvvnb {\vfifteen \noindent $\bullet$~}
\def \vvnb {\vten \noindent $\bullet$~}
\def \vnb {\vfive \noindent $\bullet$~}
\def \vn {\vfive \noindent}

\def \mb {\vspace{8pt}\nb}
\def \tb {\vspace{8pt}\nb}

\def \vvni {\vten \noindent}
\def \vni {\vthree \noindent}
\def \nb {\noindent $\bullet$~}
\def \ni {\noindent}
\def \bb {$\bullet$~}

\newcommand{\hypo}[1]{
    \begin{quote}
        \stepcounter{HYPO}{\bf Hypothesis \arabic{HYPO}:}
        {\em #1}
    \end{quote}
}

\newcommand{\taskformat}[2]{#1\textsc{#2}}

\newcommand{\task}[3]{
    \begin{quote}
    \phantomsection
    \hypertarget{task#1#2}{}
    {\bf Task \taskformat{#1}{#2}:}
    {\em #3}
    \end{quote}
}

\newcommand{\tasklink}[2]{\hyperlink{task#1#2}{\taskformat{#1}{#2}}}

\newcounter{HYPO}
\newcounter{TASK}

\newcommand{\rs}{{ResearchStaff$_1$}}
\newcommand{\pd}{{\bf Postdoc$_1$}}
\newcommand{\raOne}{{\bf RA$_1$}}
\newcommand{\raTwo}{{\bf RA$_2$}}
\newcommand{\ndv}{{\bf NDV}}
\newcommand{\ug}{{\bf Undergrad$_1$}}


\newcommand{\sssubsection}[1]{\vten\ni\textbf{\large{\textsc{#1}}}}

\newcounter{mysubcounter}
\setcounter{mysubcounter}{1}
\newcommand{\mysub}[1]{\vten\noindent\textcolor{black}{\textbf{\textbf{#1}\stepcounter{mysubcounter}}}}

\newcommand{\emptypage}{
\newpage
(empty page)
}

\newcommand{\myrotate}[1]{\begin{rotate}{90} {\bf #1} \end{rotate}}

\newcommand{\mycaption}[3]{
        \caption{
            \label{#1}
            {\bf #2. }
            {\em \small #3}
        }
}

\newcommand{\vs}{\textit{vs.}}
\newcommand{\eg}{\textit{e.g.}}
\newcommand{\ie}{\textit{i.e.}}
\newcommand{\etal}{\textit{et al.}}
\newcommand{\etc}{etc.}
\def \th {$^{th}$\xspace}

\newcommand{\sstar}{$^{*}$}
\newcommand{\stwostars}{$^{**}$}
\newcommand{\stristars}{$^{***}$}
\newcommand{\srealstar}{$^{\star}$}
\newcommand{\sdag}{$^{\dag}$}
\newcommand{\sddag}{$^{\ddag}$}

\newcounter{Xcounter}
\newcommand{\xxxreset}{\setcounter{Xcounter}{1}}
\newcommand{\xxx}{{\footnotesize\textcolor{red}{\textbf{xxx$_{\arabic{Xcounter}}$}\stepcounter{Xcounter}}~}}

\newcommand{\xxxinfig}{\textcolor{red}{\textbf{xx}}} 

\newcounter{Fcounter}
\newcommand{\freset}{\setcounter{Fcounter}{1}}

\newcommand{\finding}[1]{
\begin{spacing}{0.80}
\findingTable{#1}
\end{spacing}
}

\definecolor{fgray}{gray}{0.9}

\newcommand{\findingTable}[1]{
    \begin{table}[h!]
        \begin{tabular}{|p{3.2in}|}
            \hline
            \rowcolor{fgray}
            \findingBody{#1}\\
            \hline
        \end{tabular}
    \end{table}
    \vminten
}

\newcommand{\findingBody}[1]{{\small
        \textbf{Finding \#$\arabic{Fcounter}$:}
        \stepcounter{Fcounter}
        #1}
}

\setcounter{Fcounter}{1}

\newcounter{Icounter}
\setcounter{Icounter}{1}
\newcounter{Rcounter}
\setcounter{Rcounter}{1}

\newcommand{\myfinding}[1]{\vtwo\noindent{{\bf{Finding \#\arabic{Fcounter}:} \stepcounter{Fcounter}}#1}}
\newcommand{\myimplict}[1]{\vtwo\noindent{{\bf{Implication \#\arabic{Icounter}:} \stepcounter{Icounter}}#1}}
\newcommand{\myrecommend}[1]{\vtwo\noindent{{\bf{Recommendation \#\arabic{Rcounter}:} \stepcounter{Rcounter}}#1}}


%% file: title.tex
\def \mytitle {Dissecting CXL Memory Performance at Scale: \\Analysis, Modeling, and Optimization}

\title{{\textbf{\mytitle}}}

\author{\small Jinshu Liu, Hamid Hadian, Hanchen Xu, Daniel S. Berger\sdag, Huaicheng Li\\\small Email: \tt{\{jinshu,huaicheng\}@vt.edu}}

\date{
    \begin{tabular}{ccc}
         Virginia Tech & &
        \sdag Microsoft
    \end{tabular}
}

\maketitle

%% file: abstract.tex
\def \myabstract {%
Compute Express Link (CXL) is a promising interconnect technology that
enables system memory expansion, but it comes at the cost of long
latencies and low bandwidth compared to socket-local memory. To fully
understand the performance potential of CXL and mitigate its high
latency overhead, a detailed characterization of CXL performance
is crucial to guide the modeling and optimization of CXL
memory systems.

We present \sys, a characterization framework designed to thoroughly
analyze, model, and optimize CXL memory performance. \sys is based on 
extensive evaluation of \numWorkloads\ workloads spanning \numCxlDevs\
real CXL devices within \numLatLevels\ memory latency configurations
across \numCpus\ processor platforms. \sys\ uncovers many key insights,
including detailed workload performance at sub-\us\ memory latencies
(\cxlLatMin-\cxlLatMax\ns), CXL tail latencies, CPU tolerance to CXL
latencies, CXL performance root-cause analysis and precise
performance prediction models. In particular, \sys\ performance models
rely solely on \numPmuCounters\ CPU performance counters and accurately
fit over \modelAccEmrNuma and \modelAccEmrCXLB-\modelAccEmrCXLA workloads with a \modelTgt
misprediction target for NUMA and CXL memory, respectively.

We demonstrate the practical utility of \sys\ characterization findings,
models, and insights by applying them to popular CXL memory management
schemes, such as page interleaving and tiering policies, to identify
system inefficiencies during runtime.
We introduce a novel ``best-shot'' page interleaving
policy and a regulated page tiering policy (\alto) tailored for memory
bandwidth- and latency-sensitive workloads.
In bandwidth bound scenarios, our ``best-shot'' interleaving, guided by
our novel performance prediction model, achieves close-to optimal
scenarios by exploiting the aggregate system and CXL/NUMA memory
bandwidth.
For latency sensitive workloads, \alto, driven by our key insight of
utilizing ``amortized'' memory latency to regulate unnecessary page
migrations, achieves up to \pctTierImprv improvement over
state-of-the-art memory tiering systems like TPP, as demonstrated
through extensive evaluation with 8 real-world applications.
}

\begin{abstract}
\textit{\myabstract} \\
\end{abstract}

%% file: intro.tex
\section{Introduction}
\label{sec:intro}


The demand for increased memory capacity is rapidly rising, driven by
the growing requirements of data-intensive applications
\cite{cxlexpansion.web}.  The surge is further compounded by DRAM
scaling challenges \cite{dramscalingchallenges.imw20}.
Emerging interconnects like Compute Express Link (CXL) holds
the promise of both scale-up and scale-out coherent memory expansion at
the server/rack levels \cite{pond.asplos23, tpp.asplos23,
cxlscale.web}.
Memory vendors have introduced CXL memory expanders
\cite{cmmb.web,microncxl.web,asteracxl.web,montagecxl.web},
facilitating access to significantly larger amounts of DRAM than
previously feasible.
For instance, Samsung's CXL Memory Module - Box (CMM-B) \cite{cmmb.web}
offers 16TB of DRAM with 8 CXL devices.

\input{fig-motiv}

Memory performance is key to system performance. However, CXL memory
expansion introduces higher access latencies compared to traditional
socket-local DRAM configurations.
Figure \ref{fig:mot} illustrates the substantial heterogeneity in CXL
latency and bandwidth, as measured across various CXL devices within our
platform (Table \ref{tab:hw}) and from public
sources\footnote{\tts{CXL+Switch} data is from \cite{cmmb.web},
bandwidth averaged for 1 CXL device.}\cite{cmmb.web,skhynixcxl.web}.
Furthermore, CXL devices can exhibit varying performance
characteristics.
The variability in latency and bandwidth arises from varying
interconnection topologies and vendor optimizations.
For instance, the latencies of locally-attached CXL range from
$\sim$200-400ns, slightly exceeding cross-socket/NUMA latency. Accessing
CXL from a remote socket results in increased latency and diminished
bandwidth (\ts{CXL+NUMA}). The incorporation of a CXL switch to extend
connectivity will introduce additional latencies (\ts{CXL+Switch}), even
elevating latency to approximately 600ns.
In the future, with CXL potentially involving multiple routing hops and its use
with slow memory media (\eg, Flash) \cite{cmmh.web}, latency is projected to
increase to \us-level.

The current CPU architecture and memory hierarchy are tailored for
typical 1-2 socket systems, offering \roughly 100ns latency and 100s of
GB/s bandwidth. However, the performance implications of emerging CXL
memory technology remain uncertain.
Currently, there is a lack of research exploring detailed
CXL characteristics and its impact on memory-intensive workloads at large-scale.
Conducting a thorough characterization is crucial to provide valuable
insights for the imminent CXL deployment in production systems and software/hardware memory management.

In particular, how do CXL devices vary from each other in terms of
detailed performance characteristics? How does CXL's long latency impact
CPU efficiency and workload performance? What are the root causes?
Addressing these questions requires a deep understanding of the
dynamic nature of CXL's performance characteristics, which span a
spectrum rather than adhering to fixed, static values of latency and
bandwidth.
While previous studies \cite{thymesisflow.micro20, pond.asplos23,
cxlhpcstud.sc23, demystifycxl.micro23, cxlasicstudy.eurosys24} provide
valuable insights into CXL performance impact, they are primarily done
at a coarse-grained level, overlooking critical aspects such as CXL
performance stability (\ie, tail latencies), CPU tolerance to long
CXL latencies, CXL's architectural implications and performance
predictability.

We present {\bf \sys}, a comprehensive characterization framework for
large-scale CXL performance profiling, analysis, modeling, and
optimizations. Our goals are:

{\vni\bf (1) Understanding CXL latency and throughput implications.} How
(much) does CXL impact workload performance? What are the root causes?
And how to reason about it systematically? Can workloads benefit from
the higher aggregate memory bandwidth by splitting the dataset between
local and CXL memory and how?  We conduct a large-scale performance
study of the characteristics of \numCxlDevs CXL devices and assess
\numWorkloads workloads across \numLatLevels memory latency
configurations ranging from \cxlLatMin-\cxlLatMax\ns on \numCpus
processor platforms.
This study provides a quantitative analysis of CXL performance at scale,
uncovering new findings and insights that would not have been possible
without a large-scale approach.

{\vni\bf (2) Memory performance modeling.} Can lightweight models reliably predict
workload performance in CXL-enabled environments?
Through an in-depth root-cause analysis complementing our
characterization findings, we delve into CXL implications on CPU
efficiency and develop novel linear models for workload performance
prediction under CXL. Our models are based on novel combinations of
solely \numPmuCounters CPU performance counters but can work
surprisingly well.
We emphasize that our accurate prediction models represent a significant
advancement in enhancing the {\bf observability and predictability of memory
system performance}. They are {\it simple, easy-to-use, explainable,
general}, and can serve as fundamental performance metrics which we
believe can potentially enable many use cases.

{\vni\bf (3) Memory performance optimization.}
What are the limitations of existing memory policies in managing CXL
memory, and how can we leverage CXL characteristics to design better
memory management policies?
We show that \sys's approach can be used to quantitatively analyze the
inefficiencies of complex memory policies in managing CXL memory.
Additionally, we can leverage insights from \sys to develop enhanced
memory management strategies.
We apply \sys's characterization techniques and prediction models to
memory tiering \cite{linuxtier.web} and interleaving \cite{winter.web}.
Our experiments demonstrate the effectiveness and broad applicability of
\sys's insights in identifying system inefficiencies and enhancing the
observability of complex memory systems. More importantly, we introduce
\sys-augmented interleaving and tiering policies, which lead to
significant performance improvements
compared to state-of-the-art.
In summary, our key contributions are:

\begin{enumerate}[label=\textbf{(\arabic*)},leftmargin=*,itemsep=4pt] \vminfive
  \setlength{\parskip}{0pt}
  \setlength{\parsep}{0pt}

\item \sys, the largest-scale CXL performance study, to the
    best of our knowledge, characterizing \numWorkloads\ workloads under
    \numCxlDevs\ real CXL devices across \numLatLevels\ memory latency
    configurations on \numCpus\ processors, detailing many
    new findings about workload performance under sub-\us memory
    latencies, CXL device performance (such as latency stability) and
    deep-dive analysis of CXL and CPU interactions across workloads and
    setups.
\item A novel root-cause analysis approach based on CPU stall cycles
    for workload performance dissection under CXL, identifying
    and quantifying various sources of CXL-induced performance
    degradations in the CPU.

\item A linear performance prediction model for both latency and
    bandwidth sensitive scenarios that are {\it workload-independent and
    robust} (validated under multiple CXL and processor platforms and
    various memory policies), {\it simple and lightweight} (using
    only \numPmuCounters CPU performance counters), {\it accurate} (for
    both NUMA and CXL memory), and {\it explainable} (from root-cause
    performance breakdown analysis).

\item A ``best-shot'' page interleaving policy for bandwidth-bound
    workloads to effectively utilize both system and CXL bandwidth
    simultaneously, achieving near-ideal bandwidth
    improvements\footnote{Calculated as CXL bandwidth over system
        socket-local DRAM bandwidth, \ie,
    \tt{BW\textsubscript{CXL}}/\tt{BW\textsubscript{DRAM})}.}.

\item \alto, a memory tiering policy based on a core insight of
    ``amortized'' memory access latency by incorporating both
    memory-level-parallelism and access latencies to precisely capture
    the impact of page migrations to workload performance. By minimizing
    unnecessary page migrations and reducing the associated overhead,
    \alto achieves up to \pctTierImprv\ improvement compared to TPP
    \cite{tpp.asplos23}, a popular CXL memory tiering solution.

\end{enumerate}

%% file: fig-motiv.tex
\begin{figure}[t!]
\centerline{\includegraphics[width=.5\textwidth]{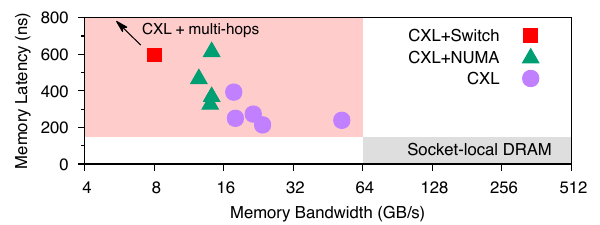}}
\vminfive
\mycaption{fig:mot}{CXL latency and bandwidth heterogeneity}{}
%
\vminten
\vminfive
\end{figure}
%

%% file: bg.tex
\section{Background and Motivation}
\label{sec:bg}

Below we present CXL background on the protocol, CPU-CXL
interactions, memory profiling, and CXL
memory management policies.

{\vthree\ni\bf CXL for memory expansion.}
CXL \cite{cxl3spec.web} is an emerging
cache coherent interconnect built atop PCIe.
It enables many potential use cases, such as memory expansion, pooling,
and sharing.
CXL memory seamlessly integrates into systems as cacheable,
byte-addressable memory within a {\it zero-core NUMA} (\znuma) node
(\ie, CPU-less NUMA) \cite{pond.asplos23}.  Thus, applications can
simply treat it as a slower-tier of memory compared to local
DRAM.
Although CXL outperforms PCIe in speed due to tailored transaction and
link protocols, it is commonly perceived that its latency is comparable
or slightly worse than that of one NUMA hop \cite{cxllatency.web}.
Moreover, CXL can increase system bandwidth, potentially benefiting
bandwidth-bound workloads.
Despite the rollout of CXL products in the last three years, there
remains a lack of in-depth studies to comprehensively understand their
performance implications, which motivates our work.

{\vthree\ni\bf CXL request processing.}
In a conventional pyramid-shaped memory hierarchy
\cite{memhierarchy.web} with L1, L2, and L3/LLC caches, if a memory request
(\eg, reading 64B of data) is not satisfied by the L1--L3 caches due to
cache misses, the request is forwarded to the CXL memory controller (MC)
via the CXL link. Once CXL memory returns the requested data,
L1--L3 caches are updated to serve future requests more efficiently. At
a high level, the CPU's request processing flow remains the same for
both local DRAM and CXL \cite{cxl-primer.csur24}. However, the use of
different buses (DIMM vs. CXL/PCIe) and MCs (on-CPU integrated, \ie, IMC
vs. third-party) affects the efficiency of CPU cache
hierarchy.

The \ts{load/store} interface is used for a CPU to communicate with
integrated or CXL MC to perform memory operations. The CPU issues two
types of \ts{load} requests: on-demand and prefetching read operations.
On-demand loads are memory read operations where the CPU requests data
from (CXL) memory only when it is needed for computation while
prefetching loads are predictive reads (directed by the hardware
prefetchers) in advance. The CPU issues store requests to write data to
memory. To maintain cache coherence, if the CPU wants to modify a
cacheline, it needs to first send a read-for-ownership (RFO) request to
gain exclusive access to the cacheline by asking the other cores to
invalidate their copies of the cacheline and/or load the cacheline from
(CXL) memory. Thus, the (CXL) MC needs to handle three types of memory
reads: on-demand, prefetching, and RFO. We will later show that
differtiating the three types of memory reads is crucial for
understanding CXL's performance implications (more in
\sec\ref{sec:rca}).

For example, CPUs heavily rely on hardware prefetchers to minimize
potential pipeline \cite{cpupipeline.web} stalls caused by the
longer access latency of (CXL) DRAM compared to L1--L3 caches. The pipeline
refers to the multiple instruction processing stages for concurrent
instruction executions, which helps improve CPU speed. However, the
increased CXL access latency can lead to delayed request prefetching,
causing the CPU pipeline to stall for a longer period (\ie, waiting for
data to arrive, more in \sec\ref{sec:rca}). This results in degraded
workload performance under CXL.

{\vni\bf CXL profiling and profile-guided optimizations.}
Modern CPUs offer robust profiling capabilities through hardware
counters/events sampling for top-down microarchitecture analysis (TMA)
\cite{tma.ispass14}. This technique has been integrated into widely used
profilers, \eg, Linux \ts{perf}. TMA allows users to pinpoint CPU
inefficiencies with well-defined metrics. For example, DRAM-bound metric
measures how often CPU was stalled on DRAM.
As modern data-intensive workloads becomes increasingly memory-bound, they can
lead to significant stalled CPU cycles
\cite{googledctax.isca15}. This approach is important for understanding
performance issues that arise from the inherent memory
access patterns of these workloads.

Leveraging such information to inform system optimizations is a
well-established practice \cite{hemem.sosp21, tpp.asplos23,
memtis.sosp23}. One common strategy involves utilizing hardware
performance counters/events, either individually or in combination,
within heuristic or ML algorithms as performance predictors. However, it
remains a challenge to define accurate performance metrics that can
capture complex system behaviors.
There are two limitations with existing approaches: accuracy and
complexity. Many widely used performance indicators, such as
\ts{LLC-miss}, are inaccurate. And ML methods introduce high
computational overhead to be useful for scenarios with tight time
constraints of 100s of ns.
Thus, TMA is mainly used for offline workload analysis. We will address this with a
clever combination of multiple performance counters to serve as reliable
performance predictors (\sec\ref{sec:pred}) and use them online for system optimizations (\sec\ref{sec:opt}).


{\vni\bf CXL memory management.}
Utilizing CXL as regular DRAM can lead to suboptimal performance due to
CXL's longer latency and/or relatively smaller bandwidth.
There are two popular approaches to address this challenge: (NUMA) page
interleaving and memory tiering. Page interleaving involves distributing
page allocations across NUMA nodes in round-robin to maximize
bandwidth usage \cite{mmpolicy.web}. In contrast, tiering aims to
minimize CXL latency impact by prioritizing local DRAM for
most-frequently accessed pages via proactive page migrations.
While interleaving and tiering have been studied across
various heterogeneous memory contexts,
including CXL, persistent memory, and disaggregated memory \cite{autonuma.web,
autotiering.atc21, hemem.sosp21, memtis.sosp23,tpp.asplos23}, fundamental
gaps remain in effective tiering policy designs.

In the rest of the paper, we present characterizations in
\sec\ref{sec:ovr} and \sec\ref{sec:rca}, CXL/NUMA performance models in
\sec\ref{sec:pred}, system optimizations for interleaving and tiering in
\sec\ref{sec:opt}, and conclude in \sec\ref{sec:conc} followed by
discussion and related work in \sec\ref{sec:dis} and \sec\ref{sec:rel}.
\bdr

%% file: ovr.tex
\section{\mbox{Overview and CXL Characterization}}
\label{sec:ovr}

\subsection{\sys Overview}
\label{s:ovr:ovr}
Figure \ref{fig:arch} provides a high-level overview of \sys pipeline.
To address the research questions raised in \sec\ref{sec:intro}, we need
to overcome the following challenges:

\input{fig-arch}

\begin{itemize2}

\item Lack of fine-grained profiling tools for {\bf in-depth} analysis
    of CXL's unique performance characteristics at {\bf request-level},
    and their impact {\bf at scale}, rather than focusing solely on
    high-level average latency and bandwidth to understand a limited set
    of workloads as in prior works \cite{demystifycxl.micro23,
    cxlasicstudy.eurosys24}.

\item Lack of {\bf systematic} approaches to analyze CXL-induced
    slowdowns and identify the {\bf root causes} of performance
    degradation, rather than treating CXL as a black box.

\item Lack of {\bf explainable} performance metrics to improve the
    observability of both 1-tier and 2-tier (with NUMA/CXL) memory
    systems, particularly under long memory latencies, rather than
    relying on heuristics.

\item Lack of {\bf deterministic} and CXL-aware data placement policies
    to exploit CXL performance potentials in memory interleaving and
    tiering setups.

\end{itemize2}

\sys introduces a suite of new benchmarking and profiling tools,
analytical and modeling approaches, findings, and memory policies to
bridge the gaps. For the first time, \sys provides a detailed analysis
of the {\it unpredictable CXL latencies and their impact on CPU efficiency}.
It aims to distill key findings applicable to a wide range of workloads
and unify them into a set of performance metrics and models using a
simple yet accurate approach based on the novelty combination of a few
CPU performance counters.
The insights derived from \sys's characterization and modeling provide
deeper understanding of how CXL's long latencies affect CPU performance.
Notably, we find that although \sys performance models are specific to
certain hardware configurations (\eg, CPU and memory), they are {\bf
independent} of the workloads, allowing them to be applied across both
offline and online scenarios.
\sys-powered memory tiering and interleaving policies not only deliver
superior performance gains but also provide valuable insights for
designing future CXL-aware memory systems.

\input{tab-cxldev}

\subsection{Platform}
\label{s:ovr:hw}

We show the details of our hardware platform in Table \ref{tab:hw}.

{\vni\bf Servers.}
We use two servers equipped with Intel's 4th (Sapphire Rapids, SPR) and
5th (Emerald Rapids, EMR) generation Xeon scalable server processors.
The two servers are identical except for
their CPUs. Each server is a dual-socket (2S) system with 16 cores per
socket, running at 2.1GHz. They are equipped with 48KB L1 data cache, 2MB L2
cache, and 8 memory channels with 128 GB of DDR5-4800MHz memory. The key
difference between them is the size of the L3/LLC cache: our EMR has a
160MB LLC, whereas SPR has only 60MB.
As a more recent processor, EMR offers better support for CXL and
delivers up to 28\% better performance than SPR for certain workloads we
measured (due to its much larger LLC).

We also use two Skylake servers -- one with 2 sockets (SKX2S) and
another with 8 sockets (SKX8S) -- to extend the range of memory
latencies from \cxlLatMin\ to \cxlLatMax ns using \znuma and by lowering the CPU
uncore frequency. Together, the setups provide a total of \numLatLevels
latency configurations (including \numCxlDevs CXL devices).
We find that the performance of \znuma and local DRAM is more stable
compared to real CXL devices, making \znuma a clean-slate environment
for our characterization and modeling (further details to follow).

{\vni\bf CXL devices.} We use \numCxlDevs CXL memory expanders from
different vendors (denoted as {\bf \cxla}, {\bf \cxlb}, {\bf
\cxlc}, {\bf \cxld}). Our CXL devices' average latency and bandwidth
are 214-394ns and 18-52GB/s, respectively, measured by Intel Memory
Latency Checker (MLC) \cite{mlc.web}.
Note that \cxld is hosted on a remote machine while others are in our
lab environment, \cxlc only supports 16GB DRAM, thus we were only able
to finish a subset of \numWorkloads workloads on them.

All our CXL devices are CXL 1.1 type-3 memory expanders (supporting
CXL.io and CXL.mem). These devices function as black boxes to us, as we
do not have access to their internal implementation details. \cxlc is
FPGA-based (lowest performance) while the rest are ASICs.
\cxld utilizes 16\tms PCIe 5 lanes and supports 4 DIMMs, providing
the highest CXL bandwidth of 52GB/s. In contrast, the other devices use
8\tms lanes and 2 DIMMs, resulting in nearly half the bandwidth
(18-24GB/s), as shown in Table \ref{tab:hw}. CXL-A and CXL-C use DDR4
memory, while CXL-B and CXL-D use DDR5 memory.
In terms of latency, interestingly, CXL-A exhibits the lowest latency at
214ns, despite using DDR4 memory, while the DDR5-based CXL-B and CXL-D
have higher latencies of 239ns and 271ns, respectively. We speculate
that these differences in performance characteristics are primarily due
to variations in CXL memory controller optimizations (\eg, scheduling
policies, row buffer management, QoS, thermal management in the
controller) \cite{memsysbook}.
Accessing CXL from a remote socket (\ts{+NUMA} column) increases the
latency and decreases bandwidth. However, to our surprise, the latency
increase via one NUMA hop vary more significantly by device, \ie,
increasing by 161ns, 202ns, 227ns, and 94ns, for CXL A--D respectively.
Later, we show CXL+NUMA leads to unexpected slowdowns for some workloads
(\sec\ref{s:char:w}) which requires careful management.

{\vni\bf Workloads.}
We use a diverse set of representative workloads for the
characterization, covering
cloud workloads (caching and DB such as Redis \cite{redis.web} and VoltDB \cite{voltdb.web}, CloudSuite \cite{cloudsuite.web}, and Phoronix
\cite{phoronix.web}),
graph processing (GAPBS \cite{gapbs.web21}, PBBS \cite{pbbs.web}),
data analytics (Spark \cite{hibench.icde10}),
ML/AI (GPT-2 \cite{gpt2.web}, MLPerf \cite{mlperf.web}, Llama
\cite{llama.web}),
and high-performance computing (SPEC CPU 2017 \cite{speccpu2017.web},
PARSEC \cite{parsec.pact08}).
Some workloads are latency-sensitive (\eg, cloud workloads), some are
bandwidth-sensitive (\eg, HPC workloads), and others are a mix of both.
We consider a large-scale study essential to uncover key findings and
insights (discussed later) that would not have been achievable with a
small-scale study.

%% file: fig-arch.tex
\begin{figure}[t!]
\centerline{
\includegraphics[width=1.01\columnwidth]{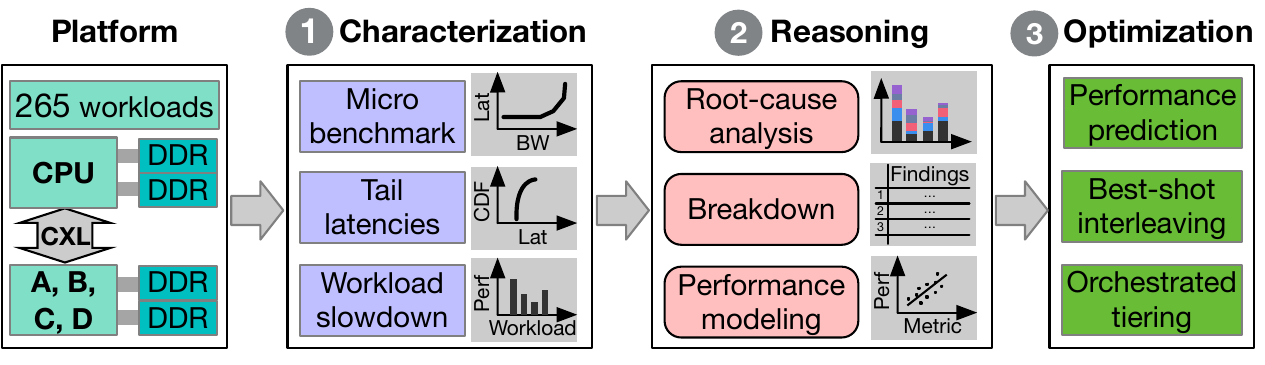}
}
\vminfive
\mycaption{fig:arch}{Overview}{Our in-depth and at-scale characterization
enable CXL performance modeling and optimization.}
\vminten
\end{figure}

%% file: tab-cxldev.tex
\setlength{\tabcolsep}{2.2pt}

\begin{table}[t!]
\begin{center}
\small
\newcommand{\ddr}[2]{\ts{#1}\texttimes\ts{DDR#2}}
\begin{tabular}{l|ll|lr|lr|r}
 & & & \multicolumn{2}{c|}{Local} & \multicolumn{2}{c|}{+NUMA} & Specification \\
\hline
    {\bf CPU} & {\bf DDR} & {\bf Size} & {\bf Lat} & {\bf BW} & {\bf Lat} & {\bf BW} & \ts{L1D-L2-L3} \\
    {\bf / CXL} & Type & GB & ns & GB/s & ns & GB/s & / \ts{CXL-dev-spec} \\
\hline
\hline
    \ts{SPR2S} & \ddr{16}{5} & 256 & 114 & 218 & 191 & 97 & 48\ts{KB}-2\ts{MB}-60\ts{MB} \\
    \ts{EMR2S} & \ddr{16}{5} & 256 & 111 & 246 & 193 & 120 & 48\ts{KB}-2\ts{MB}-160\ts{MB}\\
    \ts{SKX2S} & \ddr{16}{4} & 192 & 90 & 52 & 140 & 32 & 32\ts{KB}-1\ts{MB}-13.8\ts{MB} \\
    \ts{SKX8S} & \ddr{16}{4} & 384 & 81 & 109 & 411 & 7 & 32\ts{KB}-1\ts{MB}-38.5\ts{MB} \\
\hline
\hline
\ts{\cxla} & \ddr{2}{4} & 128 & 214 & 24 & 375 & 14 & \ts{ASIC}, \ts{CXL}1.1, {\texttimes 8} \\
\ts{\cxlb} & \ddr{2}{5} & 128 & 271 & 22 & 473 & 13 & \ts{ASIC}, \ts{CXL}1.1, {\texttimes 8} \\
\ts{\cxlc} & \ddr{2}{4} & 16 & 394 & 18 & 621 & 14 & \ts{FPGA}, \ts{CXL}1.1, {\texttimes 8} \\
\ts{\cxld} & \ddr{4}{5} & 768 & 239 & 52 & 333 & 14 & \ts{ASIC}, \ts{CXL}1.1, {\texttimes 16}
\end{tabular}
\end{center}
\vminfifteen
\mycaption{tab:hw}{Experimental platform}{``Local'' refers to the
performance measured by CPUs on the same socket while ``+NUMA''
indicates memory access from a remote socket.}

\vminten
\vminfive
\end{table}

%% file: m-char.tex
\subsection{CXL Device Characterization}
\label{s:char:dev}

We start with device-level microbenchmarks to
understand CXL latency characteristics in detail.
We run workloads using either local or CXL memory. Local DRAM
performance is used as the baseline to calculate CXL slowdowns.

\input{fig-cxl-lat-cdf}

{\vni\bf CXL latency stability and tail latencies.} To understand
latency variability of different CXL devices, we measure latencies for
each cacheline request.
As existing memory benchmarking tools do not support request-level
latency reporting, we implemented a microbenchmark program (called {\bf MIO})
that can measure cacheline-granular request latencies. MIO average
latency results are validated with Intel MLC \cite{mlc.web} reported
ones to be accurate. MIO measures the average latency of each $N$
(configurable, to amortize \ts{rdtsc} timing overhead) pointer-chasing
operations on a working set larger than LLC size. We use an in-memory
buffer from an idle NUMA node to store the latency logs to avoid
interference and minimize performance overhead.
Figure \ref{fig:cxl-cdf} shows the CXL latency distributions of all
\numCxlDevs CXL devices and Local-DRAM/NUMA under 1-32 colocated
pointer-chasing threads (from left to right).
This setup mimics the co-location of multiple memory latency-sensitive
workloads. Note that none of the CXL device bandwidth is saturated and
pointer-chasing is purely latency-sensitive operation. We disabled L1/L2
prefetchers to measure device-level latencies.

We observe \cxlb and \cxlc suffers from significantly high
tail latencies. Local and NUMA latencies are stable, and the difference
between p99.9 and p50 latencies are only 45ns and 61ns. However, CXL
latency stability largely varies across vendors.
The small latency variation for local and NUMA are probably due to DRAM
chip-level latency variations (\eg, row buffer hit/miss, activation
latencies, \etc) widely discussed in prior DRAM characterization works
\cite{cpistud.iiswc20, memanalysis.memsys20, memlatstd.iiswc15,
dramvar.sigmetrics16, demystifydram.sigmetrics19} (also in
\sec\ref{sec:rel}).
Local DRAM latency variation is much smaller than that of CXL.
For example, \cxld can deliver the best latency stability, its
difference between p99.9 and p50 is 75ns (only 30ns and 14ns more than
Local and NUMA). However, for \cxlb and \cxlc, it can reach
$\sim$160ns, which is 50\% higher than the median latency. When looking
at higher percentiles at p99.99 and p99.999, CXL device latencies will
be above 700ns for \cxla and \cxld and $>$1\us for \cxlb and
\cxlc.

Similarly, when one pointer-chasing thread is co-located with multiple
bandwidth-bound read/write threads (results not shown), we observe even
worse tail latency trends on CXL compared to Local/NUMA. When turning on
CPU prefetchers, we see effective improvement of the average latency but
tail latencies persist for CXL.

\input{fig-w-cdf}

We speculate that high CXL tail latencies are caused by the CXL
controller sub-optimal optimizations, for example, inefficiencies in
thermal management or memory request scheduling could lead to long
queueing delays. Unfortunately, there are no available tools
to investigate the exact cause of CXL tail latencies. A potential future
white-box approach could involve breaking down the latency of each
memory request and accounting for the latency across different
components, such as the CXL link, CXL controller, and DRAM chips. This
would be feasible if CXL controller exposes detailed performance
counters, for example, through the upcoming CXL performance monitoring
unit (CPMU) defined in CXL 3.0 specification \cite{cxl3spec.web},
similar to the CPU PMU. As a first step, we aim to demonstrate and
quantify the impact of CXL tail latency to raise awareness in the
systems community.
To summarize,


\myfinding{Not all CXL devices are created equal, each carrying very
unique performance characteristics. More importantly, CXL devices
exhibit unstable and higher tail latency compared to regular
socket-local or NUMA memory.
High access parallelism and high memory pressure (\eg, bandwidth)
can exacerbate CXL tail latencies.
Further, concurrent reads and writes exert differing impacts on
memory latency for CXL devices, especially regarding tail latencies.
While CPU hardware prefetchers can improve average memory access
latencies, they fail to mitigate tail latencies. CXL tail latencies
negatively impact application performance.}

\myimplict{From both software and hardware design perspectives,
there is a need to address CXL tail latencies. Future CPUs need be
improved (\eg, via smarter CXL-aware prefetching policies) to better
manage CXL's long and unpredictable latencies effectively. Additionally,
(some) CXL controllers need further optimizations to achieve latency
predictability, rather than solely focusing on average latency and
bandwidth.}

\myrecommend{Tail latency should be used as a key metric for evaluating CXL
devices, as predictable latency is crucial for meeting user service
level objectives (SLOs) in cloud environments.}

\subsection{Workload Characterization}
\label{s:char:w}

To fairly compare results from different CXL devices, we
first analyze common workloads that we
complete on all platforms followed by more workloads analysis
(\numWorkloads) on \znuma, \cxla and \cxlb.

Figure \ref{fig:wcdf}a shows the CXL slowdown CDF of 43 workloads from
SPEC CPU 2017 across 4 CXL devices on EMR and 3 \znuma
latency configurations.
The left-most black line is NUMA performance with up to 34\% slowdowns
from two bandwidth-intensive workloads (\ts{619.lbm} and
\ts{649.fotonik3d}).
Almost half of the workloads do not experience
slowdowns at all due to the large cache in EMR CPU (160MB LLC).
In total, 32 workloads experience less than 5\% slowdowns and 3 more
workloads below 10\%.
Among the four CXL devices, \cxld (green line) performs on-par with
\znuma because its high bandwidth prevents any workloads from being
bandwidth-bound.
There are four bandwith-bound workloads requiring over 24GB/s --
\ts{603.bwaves}, \ts{619.lbm}, \ts{649.fotonik3d}, \ts{654.roms} --
whose bandwidth needs exceed the capacity of CXL-\{A, B, C\}. As a
result, these workloads experience significant slowdowns (over 50\%)
compared to \znuma/\cxld, due to significant device-side queueing
delays as the CXL devices become saturated.
These four workloads see worse slowdowns under \cxlb and \cxlc.
because both the latency and bandwidth deteriorate compared to \cxla.
For the remaining workloads which do not saturate CXL bandwidth, we
observe the performance worsens with increasing CXL latency.
For example, \ts{602.gcc} slowdown goes up from 12\% up to 13\%,
21\%, and 38\% for \cxla, \cxlb, and \cxlc, respectively.
Other workloads might experience more significant performance impact
under increased latency, \eg, \ts{503.bwaves\_r} slowdown jumps from
11\% to 16\% (\cxla), 33\% (\cxlb), and 81\% (\cxlc).

\cxlc is the least performant in the four CXL devices in terms of average
latency, bandwidth, and latency stability due to the FPGA-based CXL
controller implementation. It shows significantly worse slowdown results
compared to \cxla and \cxlb. For example, \ts{649.fotonik3d} even
sees a 5.3\tms slowdown, showing a combined impact from long
(unpredictable) latency and low bandwidth.

{\vni\bf (Suspicious) CXL+NUMA performance.}
We planned to use
CXL+NUMA setup to simulate CXL memory access latency setups in the range
of 400-700ns.
However, we find workload performance under CXL+NUMA is significantly
worse even than that of 2-hop NUMA whose latency and bandwidth are both
worse, indicating issues when CXL and NUMA are used together.
In CXL+NUMA,
memory requests need to go through cross-socket interconnect (\eg, UPI)
first before reaching the CXL device.
CXL+NUMA results are shown in the ``CXL-A+NUMA'' dotted brown line in
Figure \ref{fig:wcdf}a. Surprisingly, while CXL+NUMA latency is lower
than SKX-zNUMA (375ns vs. 411ns) and bandwidth is higher (14GB/s vs.
7GB/s), CXL+NUMA performance is much worse than \cxlc, which does not
seem to make sense.
Similarly, this is true for CXL+NUMA vs. \cxlc where CXL+NUMA latency
is lower (375ns vs. 394ns). Note CXL+NUMA bandwidth is indeed lower than
\cxlc (14GB/s vs. 18GB/s), but when filtering out workloads needing more
than 10GB/s bandwidth, CXL+NUMA slowdowns are still much worse than
\cxlc.
For example, \ts{520.omnetpp} sees \mlt 5\% slowdowns under all CXL
devices, but
experiences an astonishingly high slowdown of 2.9\tms under CXL+NUMA.
Upon further analysis, we found this workload consumes \mlt 1GB/s
bandwidth (read+write), and is neither latency-sensitive or
bandwidth-sensitive. We confirm the significant slowdown is due to
much worse tail latencies under CXL+NUMA, explained next.

{\vni\bf Tail-latency impact.}
\ts{520.omnetpp} performs discrete event simulation of a large ethernet network.
In Figure \ref{fig:wcdf}b, we show the CDF of sampled memory latencies
for the workload. The plot shows little difference between Local
and \cxla (gray and blue lines), which explains the small slowdown
under
\cxla.
However, CXL+NUMA (brown line) exhibits a long tail latency starting
around p98 up to 800ns.
As we reduce the load of the workload (by reducing the number of
simulated LANs on backbone switches) to 1/2 and 1/4, we observe
consistently improved tail latencies (two dotted brown lines).
Correspondingly, the slowdown on CXL+NUMA also significantly decreases
from \roughly 290\% down to \roughly 65\% and 58\%. We believe this
serves as direct evidence that tail latencies are the root
cause of the performance slowdowns.
Similarly, 10 other workloads do not experience noticeable slowdowns
under CXL but 33\%-283\% under CXL+NUMA. These findings are consistent
for both SPR and EMR, and persist regardless of CXL device used.

{\vni\bf SPR vs. EMR.} Figure \ref{fig:wcdf}c compares the slowdowns for
SPEC workloads under SPR and EMR.
Compared to SPR, EMR features a larger LLC size and microarchitecture
optimizations for CXL, which might lead one to expect improved
performance. However, Figure \ref{fig:wcdf}c shows that the CXL
slowdowns with EMR are not significantly reduced despite the increased
LLC size, indicating that larger caches have limited effectiveness in
mitigating the impact of long CXL access latencies.
Although EMR shows slightly less slowdowns than SPR on both \cxla and
\cxlb, the CXL-induced slowdowns largely persist. This indicates that
existing caches and/or prefetchers are not effective at hiding long
memory latencies. These findings suggest that simply increasing CPU
cache size is insufficient for optimizing CXL. Future CPU designs will
need to incorporate further optimizations to better mitigate the impact
of CXL's long latencies.

{\vni\bf All workloads.} Figure \ref{fig:wcdf}d presents the
slowdown CDF for \numWorkloads workloads on both EMR/SPR and
\cxla/\cxlb.  Compared to the CPU 2017 results in Figure
\ref{fig:wcdf}c, the slowdowns are more prounced as workloads from other
benchmarking suites, such as graph and ML/AI, tend to be
memory-intensive, leading to greater performance degradations. However,
the overall performance patterns remain consistent, \eg, EMR outperforms
SPR (albeit by a small margin) on both CXL devices.
On EMR, more than 15\% of the workloads experience over 50\% degradation
on \cxla, while this percentage increases to 20\% for \cxlb due to
its higher latency (and/or less predictable latency).  For SPR, 16\% and
22\% of workloads exhibit over 50\% performance degradation on \cxla
and \cxlb, respectively.
The slowdown CDFs also reveal a clear ``tail,'' with 5\% of the
workloads suffering from slowdowns of 2.3-6.3\tms, primarily due to
being bandwidth-bound.

In summary, the key takeaways from the workload-level charcterizations
are as follows:

\myfinding{%
\begin{itemize2}
\item Workload performance deterioates superlinearly with increasing CXL
    latency; more importantly, the relative slowdowns exceed the rate of
    the latency increases).
\item Longer CXL latencies correspond to worse bandwidth (CXL
    A$\rightarrow$B$\rightarrow$C), which has a more pronounced impact
    on bandwidth-bound workloads than purely latency-sensitive workloads
    due to the combined effects of increased latency and limited
    bandwith.
\item CXL devices with worse tail latencies (\eg, \cxlb and \cxlc)
    experience more significant slowdowns across all evaluated
    workloads.
\item On a positive note, many workloads can tolerate long CXL latencies
    (up to 410ns) and thus experience minimal slowdowns, suggesting that
    CXL could be useful for real-world applications in pooling
    scenarios.
\end{itemize2}
}

\myimplict{As future CXL devices are expected to significantly increase
bandwidth (\cxld is a good example, and bandwidth can also be
easily enhanced through hardware interleaving across multiple CXL
devices) and moderately reduce latency, we anticipate that future
CXL workload slowdowns will be smaller than those shown in Figure
\ref{fig:wcdf}a. Higher CXL bandwidth will benefit
bandwidth-bound workloads, potentially alleviating the 2-6\tms\
slowdowns observed in Figure \ref{fig:wcdf}a due to the low
bandwidth of individual CXL devices. Reductions in latency will
improve the performance of latency-sensitive workloads, such as
cloud applications, bringing it closer to NUMA performance.}

\myrecommend{CXL latency is more critical to performance when bandwidth
is no longer a bottleneck (see Figure \ref{fig:wcdf}a) and deseves
more attention in future CPU/CXL designs as well as software
optimizations.
However, for bandwidth-bound workloads to effectively utilize the
combined bandwidth of local and CXL memory, improved software approaches
are still needed.}

%% file: fig-cxl-lat-cdf.tex
\begin{figure}[t!]
\centerline{
\includegraphics[width=.47\textwidth]{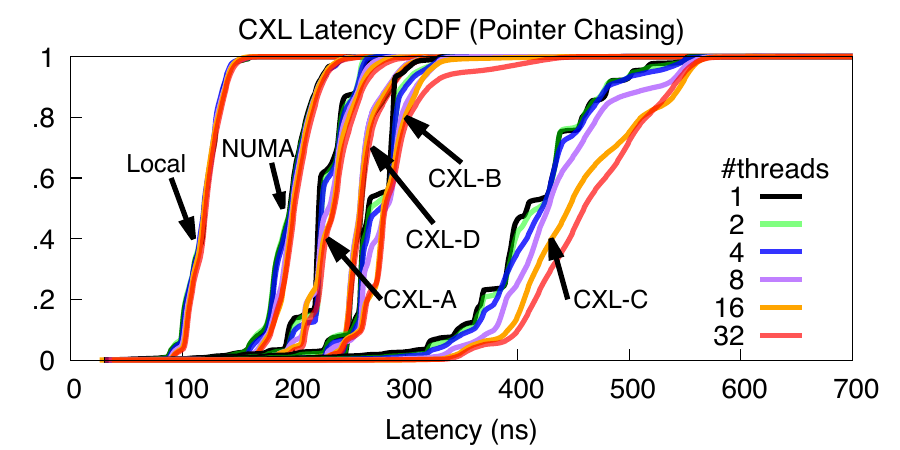}
}
\vminfive
\mycaption{fig:cxl-cdf}{CXL Latency CDF}{Not all CXL are created equal.
Unlike local/NUMA memory, CXL shows high tail latencies.}
\vminten
\end{figure}

%% file: fig-w-cdf.tex
\begin{figure*}[h]
\centerline{
\includegraphics[width=.58\columnwidth]{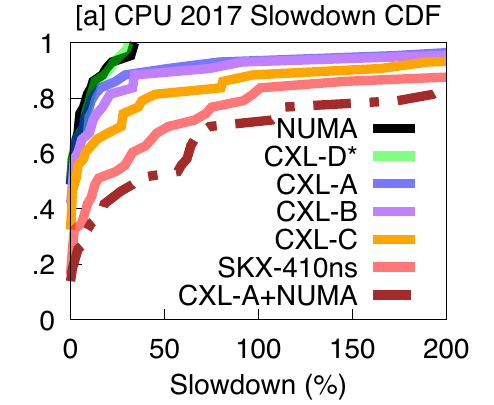}
\hspace*{-16pt}
\includegraphics[width=.58\columnwidth]{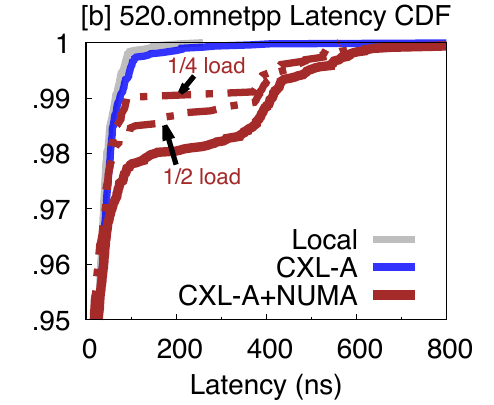}
\hspace*{-16pt}
\includegraphics[width=.58\columnwidth]{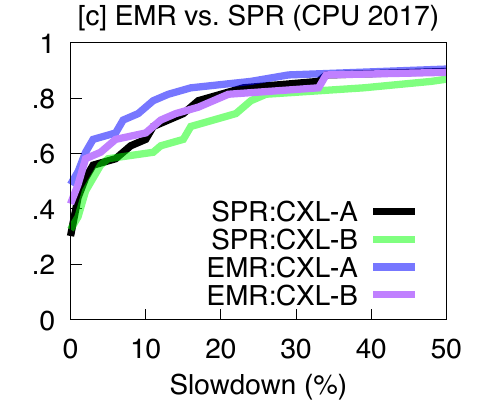}
\hspace*{-16pt}
\includegraphics[width=.58\columnwidth]{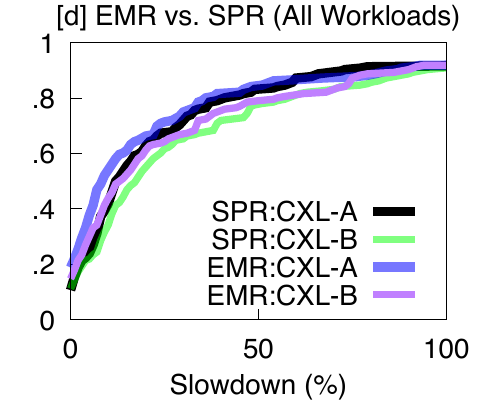}
}
\vminfive
\mycaption{fig:wcdf}{CDFs of workload slowdowns under various CXL}{(a) the
CDFs of SPEC workloads on all our platforms; (b) tail latency is the
cause of significant workload slowdown under CXL+NUMA for a
latency-insensitive workload; (c) SPR vs. EMR SPEC results under
\cxla and \cxlb; (d) is similar to (c) but for all \numWorkloads
workloads.}
\vminten
\end{figure*}

%% file: m-brk.tex
\section{Performance Modeling} 
\label{sec:rca}

\subsection{Slowdown Root-Cause Analysis}
\label{s:rca:method}

Our goal is to break down workload slowdowns into contributions from the
CPU cache hierarchy and CXL memory.
We aim to quantify the impact of each component to better understand how
CXL affects CPU efficiency. For example, instead of the general notion
that CPU prefetchers become less effective under CXL's longer latencies
\cite{againstcxl.hotnets23}, we will {\it measure} CXL's impact on
prefetcher performance and {\it disclose} why it happens.

To achieve this, we need an approach to {\it capture} the events in the
CPU pipeline that lead to performance slowdowns under CXL and correlate
them accurately back to workload-level slowdowns.
The extensive microarchitecture-level information offered by CPU PMU
counters provides valuable insights into the efficiency of the CPU
pipeline.
While workload slowdowns can be directly measured using
application-level metrics, identifying the underlying PMU events/metrics
that can correlate to the slowdowns is often challenging.
It is even more challenging to establish a precise correlation between
workload performance and architecture-level performance metrics.
The Intel TMA method \cite{tma.ispass14} is a popular approach for
top-down performance analysis, but it is insufficient for our
objectives.

\begin{enumerate2}

\item TMA identifies dominant performance bottlenecks in an application
    by analyzing execution inefficiencies within the CPU pipeline for a
    fixed setup using either local DRAM or CXL memory. However, it does
    not provide a differential analysis to interpret pipeline differences
    resulting from varying backend memory.

\item Although a differential analysis can be done manually, there is
    {\it no method to precisely correlate
    microarchitecture level metrics with workload slowdowns}. The TMA
    metrics are designed to capture the performance or contention of
    specific hardware components rather than overall workload behavior.

\end{enumerate2}

\input{fig-uarch}

For these reasons, we begin by examining components of the CPU pipeline
involved in instruction execution and analyzing the {\bf changes}
induced by CXL on those components during memory request processing. As
discussed in \sec\ref{sec:bg}, processing CXL memory requests requires
traversing the memory hierarchy, including L1, L2, LLC,
and CXL memory.
By evaluating the CPU's efficiency at these key points, we can identify
the corresponding slowdowns caused by CXL across workloads.
Figure \ref{fig:uarch}a highlights the key components as observation
points for memory request processing during CPU and CXL interactions.
Through detailed offline analysis, we make a few key observations that
lead to an accurate slowdown breakdown method which we describe below.

Workload performance slowdowns can be represented using
microarchitecture-level performance counters and reasoned about by
checking where ``stalls'' happen in the CPU pipeline.
For example, if a workload takes $c$ cycles to complete on local memory
and $c^{\prime}$ on CXL, the slowdown can be denoted as
$S=\frac{c^{\prime} - c}{c}=\frac{\Delta_{c}}{c}$.

\myfinding{The variance in cycle counts between CXL and local DRAM
primarily stems from {\it stall cycles} difference, which further mainly
arises from the CPU pipeline {\it backend}.}

As such, CXL slowdowns can be estimated as:

\begin{equation}
    S = \frac{\Delta_c}{c} \approx \frac{\Delta_{stall}}{c} \approx \frac{\Delta_{backend-stall}}{c}
    \label{eq:ss}
\end{equation}

CPU backend refers to memory-subsystem. Purely CPU-bound workloads are
not sensitive to CXL latency due to few CXL accesses, thus experiencing
minimal slowdowns.

{\vni\bf Accuracy.}
To validate the finding, we measure (backend) stall cycles for each
workload and use them to estimate the workload slowdowns according to
\eqref{eq:ss}. We compare them with the actually observed workload
slowdowns using application-level metrics (\eg, time, throughput).
Figure \ref{fig:brk-acc} presents the CDF plots of the absolute
difference between the actual slowdown and the (backend) stall based
slowdown estimations, which indicates the inaccuracies. We show the
results for \znuma, \cxla, and \cxlb.
We observe very low inaccuracies -- within 5\% for over 95\% of
workloads (the intersection of two gray lines). Therefore, CXL-induced
(backend) stall cycle difference can effectively represent the slowdown.

\input{fig-brk-acc}

\myimplict{Workload slowdowns on CXL are primarily due to the additional
backend stalls, which are caused by memory subsystem inefficiencies.}

{\vni\bf Reasoning.}
The CPU pipeline is divided into two parts: the frontend and the
backend. In the frontend, instructions are fetched and decoded, while in
the backend, they are executed. Stalled cycles can occur due to stalls
in either the frontend, the backend, or both. However, frontend stalls
are negligible because modern CPU instruction caches are efficient and
large enough to fetch and decode instructions without being affected by
CXL delays. Therefore, it is primarily stalls in the memory subsystem
(\ie, the CPU backend) that are impacted by CXL. As a result, stalled
cycles in the memory subsystem can serve as a suitable approximation for
slowdown caused by CXL.

{\vni\bf Breaking down the slowdown.}
Figure \ref{fig:uarch}a highlights the simplified CPU backend
components where the majority of these stall cycles occur, including
the {\it store buffer} for serving writes, {\it L1--LLC}, and {\it CXL}
for serving reads. By observing the number of stall cycles on each
component, we can further understand how (much) each of these backend
components contribute to workload slowdowns.

On Intel platforms, the stalls on the store buffer, L1, L2, LLC, and
(CXL) DRAM represent exclusive events which sum up to the total backend
stall cycles (see Figure 4 in \cite{tma.ispass14}). Let $s$ be the number
of stall cycles, according to TMA approach, we have:
\vspace{-.5em}
\begin{align}
    s_{Local} = s_{store} + s_{L1} + s_{L2} + s_{L3} + s_{DRAM} \\
    s_{CXL} = s^{\prime}_{store} + s^{\prime}_{L1} + s^{\prime}_{L2} + s^{\prime}_{L3} + s^{\prime}_{DRAM}
\end{align}

In the above formula, $s_{L1}$ and $s^{\prime}_{L1}$ denote the number
of stall cycles on local and CXL memory, respectively, due to L1 cache
accesses. Other terms follow a similar definition.
When looking at the difference between the two, we get:

\vspace{-1.5em}
\begin{align}
\hspace*{-16pt}\Delta_{stall} &= s_{CXL} - s_{Local} = \nonumber \\
&\hspace*{-16pt}\Delta s_{store} + \Delta s_{L1} + \Delta s_{L2} + \Delta s_{L3} + \Delta s_{DRAM}
\end{align}

Here, $\Delta s_{L1} $ denotes the difference ($\Delta$) of stall cycles
on L1 on local and CXL DRAM. Correspondingly, by dividing each item with
total cycle-count ($c$), the overall slowdown can be represented as the
combined slowdowns from the five sources as follows:

\vspace{-1em}
\begin{equation}
    S \approx S_{store} + S_{L1} + S_{L2} + S_{L3} + S_{DRAM} 
\end{equation}

Above, each component-wise slowdown is calculated as the delta of stall
cycles on the specific component, \eg, slowdown due to L1 cache access
is $\Delta$ of stalled cycles on L1, denominated by the total cycle
count ($c$), \ie, $S_{L1}=\Delta s_{L1}/{c}$.

\input{m-rc} 

Next, we will apply this approach to various workloads. Our aims are
twofold: validate the plausibility of our assumptions; and illustrate
how the breakdown method can reveal interesting insights overlooked in
prior research.

\subsection{Workload Slowdown Diversity}
\label{sec:rca:rst}

Figure \ref{fig:brk} depicts the overall and breakdown of CXL slowdowns
for each workload under \znuma, \cxla and \cxlb.
``Other'' indicates the slowdown contribution which is not captured via
our analysis.
The breakdown allows us to further analyze various causes of CXL
slowdowns. Below we summarize some findings.

For different workloads, the contribution of slowdown from various
sources varies. Taking SPEC workloads such as \ts{519.lbm},
as an example, the majority of the
slowdown originates from stalls in the CPU's store buffer. This indicates a
high volume of RFOs and insufficient entries in the CPU's store buffer.
These observations are further supported by observations such as high
UPI non-data traffic
and high write bandwidth.
However, in workloads like \ts{649.fotonik3d}, a significant portion of the slowdown arises from the
cache.

For GAPBS workloads, the primary source of slowdown is from DRAM (stalls
in LLC miss demand reads). Only a few, such as \ts{bc-urand},
\ts{sssp-web}, and \ts{bfs-urand}, encounter slowdown from the cache.
Many of the Llama workloads experience L3/LLC slowdowns.
Cloud workloads such as Redis and VoltDB, mainly suffer from DRAM
slowdowns. Similarly, DRAM slowdowns take up 90\% of the overall
slowdowns for ML workloads like DLRM and GPT-2.

\input{fig-brk-cdf}

Figure \ref{fig:brk-cdf} shows the CDFs of slowdowns caused by various
components. Briefly, at least 15\% workloads experience at least 5\%
cache slowdown on CXL, indicating the degraded prefetch efficiency under
CXL. Meanwhile, at least 40\% workloads experience with at least 5\%
DRAM slowdown.
Interestingly, L2 cache slowdown prevails as the dominant factor across
all examined workloads in the breakdown analysis (on SKX-\znuma).
Notably, deteriorated memory latency and decreased memory bandwidth
contribute to an upsurge in stalled cycles in the L2 cache.
Additionally, the stalled cycles in L1 and L3 remain relatively
unaffected.

Certain workloads, such as \ts{627.cam4}, \ts{607.cactusBSSN}, and
\ts{602.gcc}, demonstrate similar CXL performance slowdowns. However,
the reasons behind the performance slowdowns vary significantly among
them. In \ts{602.gcc}, half of the slowdown stems from LLC misses, while
the other half arises from cache. Conversely, almost all slowdown in
\ts{607.cactusBSSN} results from LLC misses, while for \ts{627.cam4},
reads caused by stores (RFOs) dominate the performance slowdown. This
underscores one of the advantages of the breakdown method, as it
highlights that although the performance slowdowns may appear similar,
the underlying causes can be vastly different.

To summarize, our approach could capture, explain and breakdown CXL
slowdowns based on the CPU stall cycles approach.
Later, we will further enhance our approach for CXL performance
prediction to show its efficacy.

%% file: fig-uarch.tex
\begin{figure}[t!]
\centerline{
\includegraphics[width=\columnwidth]{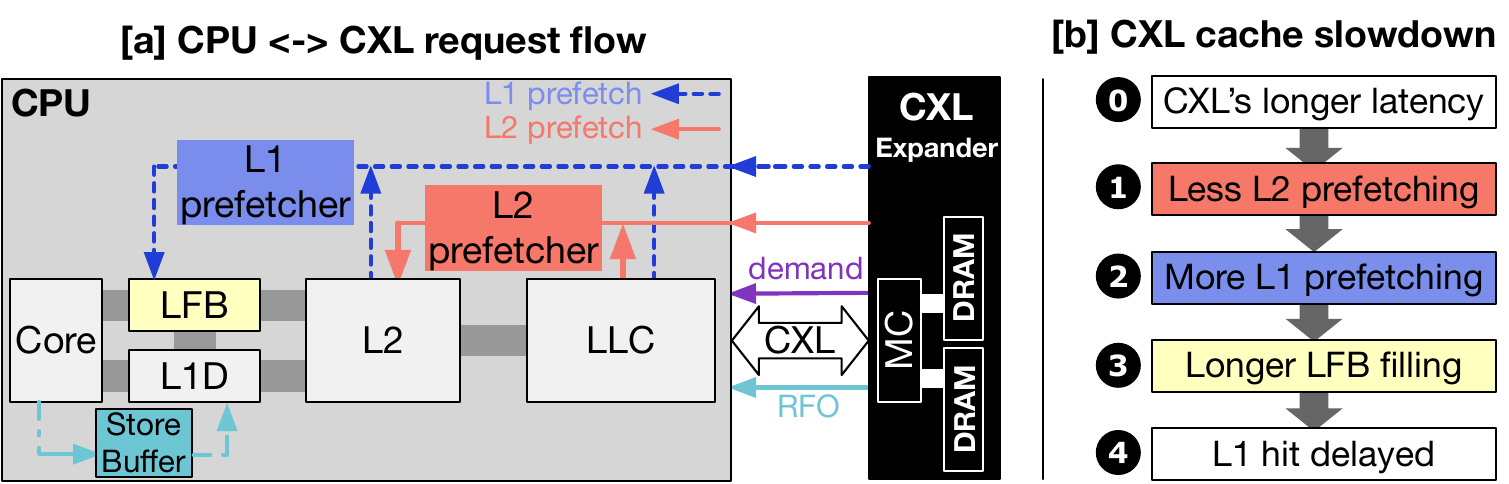}
}
\vminfive
\mycaption{fig:uarch}{CXL slowdown breakdown}{Figure (a) shows various
components where CXL introduces overheads; Figure (b) details the flow
of CXL-induced cache slowdowns.}
\vminten
\end{figure}

%% file: fig-brk-acc.tex
\begin{figure}[t!]
    \centering
    \begin{subfigure}[t]{0.35\linewidth}
        	\centering
        	\includegraphics[width=\textwidth]{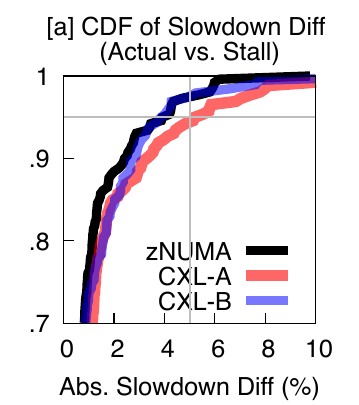}
        	\vspace{-5mm}
            \label{fig:brk-acc-stall}
    \end{subfigure}
    \begin{subfigure}[t]{0.35\linewidth}
        	\centering
        	\includegraphics[width=\textwidth]{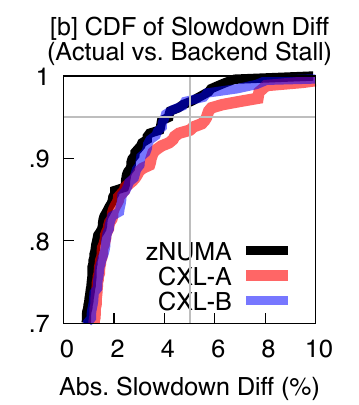}
        	\vspace{-5mm}
            \label{fig:brk-acc-backend}
    \end{subfigure}
    \vminfive
    \mycaption{fig:brk-acc}{CDFs of slowdown differences using stalls and
    backend stalls}{The X-axis represents the absolute difference between
    estimated slowdowns using stalls or backend stalls and the actual
    measured slowdowns for each workload.
    }
    \vminten
\end{figure}

%% file: m-rc.tex
{\vni\bf DRAM (Demand Load) Slowdown ($S_{DRAM}$).}
%
We use the increase in stalled cycles of LLC misses,
as a primary indicator of CXL slowdown from DRAM. These misses denote
{\it demand read misses}, excluding RFO and prefetch requests.
On Intel platforms, they are characterized as cycles stalled while LLC
demand read misses are unresolved.
Hence, their change suggests performance deterioration originating from
DRAM, including the (CXL) memory controller.
We also identify memory level parallelism (MLP) as another key metric
for analyzing slowdowns. Later, we will show how it enhances slowdown
prediction in \sec\ref{sec:pred:dram}.

{\vni\bf Store Slowdown ($S_{store}$).}
We use the increase of cycles bound on full store buffer to gauge store
operation slowdown. Incoming \ts{store} requests queued in the store buffer are
dequeued upon completion. Some writes issue RFO requests before
execution. If the store buffer fills up, these RFOs would hinder load
efficiency, causing CPU stalls.

\subsection{Cache Slowdown ($S_{cache}$)}
\label{sec:rca:cache}

While DRAM and store slowdowns are relatively straightforward to
understand, cache slowdowns are more complex. In this section, we
discuss our key findings on how CXL can degrade CPU cache efficiency.
Cache slowdown ($S_{L1}+S_{L2}+S_{L3}$) indicates stall cycle increase
on various cache levels (L1, L2, and LLC). Similarly, they can be
measured using the corresponding stall cycles counters. Below we
describe our findings to reason about cache slowdowns on CXL through
offline analysis.

\myfinding{
\begin{enumerate2}

\item Cache slowdown under CXL is due to reduced prefetch efficiency.
    To validate this, we disable all the hardware prefetchers (L1 and L2,
    LLC-prefetcher is disabled by default) and measure workload
    slowdowns. With prefetchers off, we found virtually no stall cycles
    on cache ($S_{L1}=S_{L2}=S_{L3}=0$).

\item Through our extensive offline analysis, we find CXL's relatively
    longer latency causes L2-prefetcher inefficiency (less useful data
    in L2 cache), thus causing L1-prefetcher to fetch more data from
    LLC/CXL. As a result, L1 demand reads are affected negatively (more
    stalls in L1), thus causing cache-slowdown.

\item Upon further analysis, we find cache slowdown is mainly reflected
    as the increase of hits on line fill buffer (LFB), a per-core small
    buffer with 10-20 entries that connects L1 and L2 caches. Due to the
    reduced L2 prefetcher efficiency, L1 prefetcher fetches more data
    from LFB, causing higher LFB hits.

\end{enumerate2}
}

To summarize, as shown in Figure \ref{fig:uarch}b, CXL initially leads
to reduced efficiency of L2 prefetchers. With less useful data in L2
cache, L1 prefetchers are compelled to fetch more data from LLC or (CXL)
DRAM due to L2 misses. Moreover, CXL affects L1 prefetch efficiency
as well. Data fetched by L1 prefetchers must be temporarily stored in
LFB before reaching L1 cache, and this would cause more requests to be
served by (slower) LFB hits instead of direct L1 hits, causing L1
slowdowns.

\input{fig-pref-skx}

\input{fig-brk} 

{\vni\bf \mbox{Reasoning of reduced L2-prefetch efficiency under CXL.}}
Through offline analysis of cache-related PMU counters for
local-DRAM and CXL, we find reduced number of L2 prefetch requests
that misses L3/LLC (L2-prefetch-L3-miss) on CXL. Meanwhile, L1
prefetch that misses L3/LLC (L1-prefetch-L3-miss) increases. The
increase is almost the same as the decrease of L2-prefetch-L3-miss, as
shown in Figure \ref{fig:pref-skx}a, while L2-prefetch-L3-hit does not
change.  The decrease of L2-prefetch-L3-miss on CXL setup indicates the
L2-prefetcher fails to fetch as much data as on local-DRAM-setup from
CXL, thus reducing L2-prefetcher efficiency. As a result, the
L1-prefetcher can't find data from L2 cache that should be fetched by
L2-prefetcher and it has to fetch more data from CXL, which explains the
increase in L1-prefetch-L3-miss. Figure \ref{fig:pref-skx}a shows that
the decrease of L2-prefetch-L3-miss has a strong positive relationship
with the increase of L1-prefetch-L3-miss (almost $y=x$), with a Pearson
coefficient of 0.99.

{\vni\bf Cache slowdown can be observed via LFB-hit increases.}
LFB connects L1 and L2 caches. The data of all read requests must be
placed in the LFB before reaching L1 cache from L2 or lower levels, as
in Figure \ref{fig:uarch}a. Due to its limited size, LFB can become a
bottleneck for data flowing to L1 cache. For example, Figure
\ref{fig:pref-skx}b shows that the increase in L1-stalled-cycles
correlate with high pressure on LFB (more LFB hits), caused by
L1-prefetch-L3-miss increasing.
Particularly, the increase in LFB hits (difference between CXL and
Local-DRAM) is (almost) {\it linearly} correlated with the increase in
L1-prefetch-L3-miss.
It means that more data is fetched from CXL to L1 cache by the
L1-prefetcher, which becomes LFB hits.

Similarly, the increase in LFB hits is positively correlated with the
decrease in (demand read) L1 cache hits. The reason
is that the data fetched by L1-prefetcher first goes to LFB, but has not
yet been transferred to L1 cache, due to the longer memory latency of
CXL. The data required by load instruction is fed by LFB but not L1
cache, resulting in L1 hit becoming delayed hit on LFB.

In summary, if a workload heavily relies on data from L1 prefetch (\eg,
sequential, stride, or streaming access), and this data primarily
originates from DRAM, with subsequent data often in the same cacheline,
then the stall cycles of L1 demand misses may worsen. Consequently, such
workloads are prone to experiencing high L2 cache slowdown under CXL. We
also observed that on SPR/EMR, cache slowdown predominantly arises from
LLC rather than L2 (SKX+\znuma), validated similarly.

%% file: fig-pref-skx.tex
\begin{figure}[t!]
\centerline{
\includegraphics[width=.8\columnwidth]{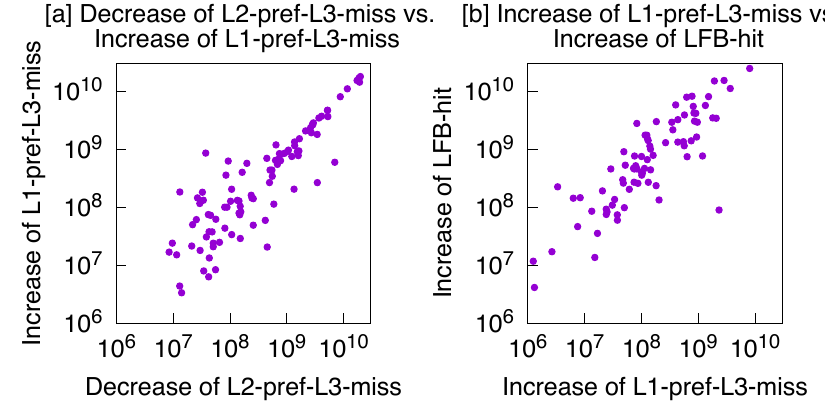}
}
\vminfive
\mycaption{fig:pref-skx}{Correlations of LFB-hit, L1-prefetch-L3-miss
and L2-prefetch-L3-miss}{(a) shows strong linear correlations of L2
prefetches that miss L3 and increase of L1 prefetches that miss L3; (b)
shows a similar trend for increase of L1 prefetches that miss L3 and
increase of LFB hits.}
\vminten
\end{figure}

%% file: fig-brk.tex
\begin{figure*}[t!] \centerline{
    \includegraphics[width=\textwidth]{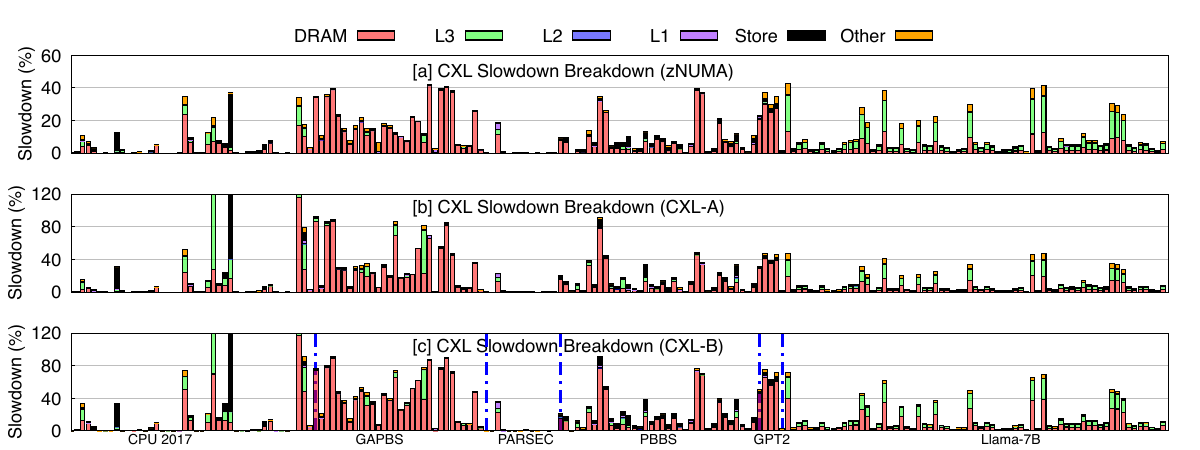} }
    \vminfive
    \mycaption{fig:brk}{CXL slowdown breakdown}{This figure shows the
    CXL slowdown breakdown on \znuma, \cxla, and \cxlb.}

%
\vminten
\end{figure*}

%% file: fig-brk-cdf.tex
\begin{floatingfigure}[r]{1.5in}
    \vspace{-0mm}
    \centering
    \hspace*{-8pt}\includegraphics[width=1.7in]{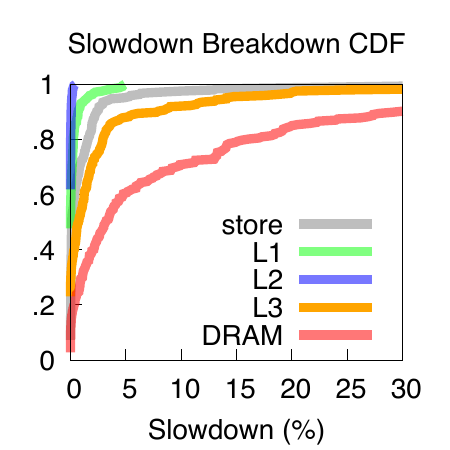}
    \vspace{-4.5mm}
    \vminfive
    \mycaption{fig:brk-cdf}{CDFs of slowdown breakdown}{}
\end{floatingfigure}

%% file: m-pred.tex

\section{CXL Slowdown Prediction}
\label{sec:pred}

The capability to predict system performance is appealing due to its
wide range of applications. Our previous root-cause analysis of CXL
slowdowns has helped identify various sources of slowdown, which, when
combined, can facilitate reasoning about measured CXL performance. In
this section, our objective is to transition and solidify our breakdown
analysis into formal prediction models. In particular, when the model is
used together with an offline workload run on local DRAM, it can
accurately predict the amount of slowdowns when the workload runs on
CXL. Later, we will also show the prediction model can be used in
an online fashion for performance optimations.

\subsection{Strawman}

We initially explore simple correlations between commonly used
performance metrics such as LLC miss rate, represented as
misses-per-kilo-instructions, (MPKI, Figure \ref{fig:ncor}a), read
memory bandwidth (Figure \ref{fig:ncor}b), and TMA DRAM-bound metric
(Figure \ref{fig:ncor}c), as they are used in many prior works
\cite{demystifycxl.micro23, pond.asplos23}. However, none of these
metrics prove reliable as performance predictors.
For example, despite a positive relationship between read bandwidth and
the overall slowdown, read bandwidth falls short as a reliable
predictor. Workloads with similar bandwidth often experience varying CXL
slowdowns, \eg, 5-50\% under 10-20GB/s.
We attribute this to the limitations of the aforementioned metrics in
capturing CXL slowdowns across diverse sources. This prompts us to
develop separate prediction models for cache, DRAM, and store-induced
slowdowns. These efforts result in several simple models, which can be
combined to accurately predict overall CXL slowdowns, relying solely on
\numPmuCounters counters on SPR/EMR (11 on SKX).

\input{fig-pred}

\subsection{Latency and Bandwidth Sensitivity}
\label{sec:opt:sensitivity}

Workload performance is influenced by both memory latency and bandwidth.
Bandwidth-sensitive workloads can benefit from increased memory
bandwidth through technologies like CXL, while latency-sensitive
workloads are better managed with tiering strategies to mitigate latency
impacts.
Therefore, accurately determining a workload's sensitivity to bandwidth
or latency is crucial.

We propose using a CPU offcore latency-based model for this purpose. Our
benchmarking results indicate that under bandwidth
contention, queueing delays contribute to end-to-end request latencies.
Offcore latency reflects both memory latency and
bandwidth-induced overhead. A simple heuristic is to set an offcore
latency threshold. If latency exceeds this threshold, it indicates
bandwidth limitation; otherwise, it is latency-bound. Additionally, the
offcore latency threshold can be easily profiled using pointer-chasing style workload,
as in \sys tail latency analysis.

We used this approach to filter out bandwidth-bound workloads on CXL,
which experience much higher slowdowns (\ie, the tail in Figure \ref{fig:wcdf}d)
where slowdowns can be up to 6\tms.

\subsection{DRAM (Load) Slowdown Model}
\label{sec:pred:dram}

There are two insights in our DRAM slowdown prediction.
The first is the overall ratio of stalled cycled on LLC (\ie,
``$P_4$ / $P_1$'') as a base predictor can already positively
correlate with DRAM-sourced slowdown. We started this analysis on SKX2S
\znuma.
In Figure \ref{fig:pm-dram}a, we correlate the based predictor observed
when the workloads run under local DRAM (90ns) with the DRAM-slowdown in
\znuma (140ns).
Notably, the predictor does a great job for most workloads showing a
strong linear relationship, with a few outliers on the top right, indicating the
predictor is mistakenly overpredicting the slowdowns.

\input{fig-cxl-mlp}

Second, we argue that not taking high {\bf memory-level parallelism (MLP)}, more precisely,
overlapping effect, into account is the cause of the above outliers.
As shown in Figure \ref{fig:mlp}, CXL has the same impact on each single data request.
For each data request, the latency will be increased similarly, \eg, $x$.
However, under high MLP,
the overlapping effect lowers the CXL impact on the slowdown (DRAM load),
as in Figure \ref{fig:mlp} left, reducing the latency from $x/a$ to $x/(a+b)$
($b$ indicates the stalled cycles from previous demand requests caused by overlapping).
A large amount of demand reads could cause considerable LLC miss stalls,
but the increase of stalled cycles of previous demand read misses could be overlapped by the last several demand read misses.
The increased LLC miss stalls of part of the data requests
impacted from low memory latency efficiency could be overlapped by the other demand reads.
In contrast, if the demand reads are spaced out,
more demand reads could be affected by memory latency and further influence
the overall increase of LLC stalls caused by remote (CXL) memory.
Therefore, we assume that the degree of overlapping would decrease the CXL impact on the (DRAM) slowdown.

Unfortunately, this effect cannot be directly measured. Instead, we choose to
approximate it using the amortized offcore demand read latency.
By incorporating MLP into the model, Figure \ref{fig:pm-dram}d shows a
much stronger linear relationship (Pearson coefficient goes up from 0.905 to 0.965).

{\vni\bf Accuracy on SPR/EMR with real CXL.}
Figure \ref{fig:pm-dram}b-c\&e-f show the DRAM slowdown models for
\cxla and \cxlb. Similar to \znuma, it could predict the DRAM
slowdown reliably.
Applying MLP impact to the model still helps improve model accuracy on
SPR/EMR, but less so compared to SKX. We speculate this is because latest
EMR CPUs with large LLC cache experiences less MLP, thus less outliers
caused by it.
Table \ref{tab:dram-err} shows the store slowdowns of \textbf{92.0\%},
\textbf{94.0\%} and \textbf{78.7\%} workloads can be predicted within
\textbf{5\%} deviation on \znuma, \cxla and \cxlb, respectively.

\input{tab-model-dram-err}

\subsection{Cache (Load) Slowdown Model}

\input{fig-pred-cache-store}

Cache introduced slowdowns are hard to directly measure and quantify.
We develop a metric to predict cache slowdowns based on our root cause
analysis. Workloads spending more stalled cycles on L2 cache, accessing
increased data on LFB, allocated by L1 prefetching requests missing on
L3, and primarily prefetched from DRAM by L1 prefetchers, may encounter
elevated cache slowdowns. Leveraging pertinent performance counters, our
predictor aims to effectively capture the contention indicative of cache
slowdowns.

Intel offers counters for helping derive performance predictor ($M_{cache}$) for
cache slowdown ($S_{cache}$).

\input{tab-model-cache-err}

Table \ref{tab:cache-err} shows the store slowdowns of \textbf{94.7\%},
\textbf{83.2\%} and \textbf{83.6\%} workloads can be predicted within
\textbf{5\%} deviation on \znuma, \cxla and \cxlb, respectively.

\subsection{Store Slowdown Model} 

We find that bound-on-store counter is positively related to store
slowdown.
It means that on remote memory (CXL),
it is increased by the same factor for most
of the workloads.

\input{tab-model-store-err}

Table \ref{tab:store-err} shows the store slowdowns of \textbf{97.8\%},
\textbf{87.6\%} and \textbf{91.4\%} workloads can be predicted correctly
within \textbf{2\%} deviation on \znuma, \cxla and \cxlb,
respectively.

\subsection{Put It All Together}
\label{sec:pred:all}

Each component contributing to the breakdown of slowdown can be
individually predicted by our introduced model. The overall slowdown is
determined by summing the slowdown from DRAM, cache, and store. Given
that most real-world servers share similar architectural organizations,
such as multiple cache levels, a store buffer, LFB, SQ, and L1/L2
prefetching, we believe this methodology can be universally applied
across different server models to analyze and predict performance
slowdowns caused by sub-\us memory latencies.

The limited availability of CPU counters impacts the accuracy of
performance modeling under CXL. Nevertheless, we demonstrate
that by meticulously integrating multiple counters in a novel manner, we
can effectively capture system performance and use it for reliable
performance prediction.

\input{fig-model-eq}

{\vni\bf Mispredictions.} Mispredictions may arise partly due to the
absence of certain performance counters provided by Intel. First,
measuring the proportion of L1 prefetching data requests within LFB hits
is impractical. Second, gauging L1 prefetching hits on L2 cache, even
with the total number of prefetching data requests from LFB hits known,
remains unfeasible. Therefore, we solely employ the L1 prefetching L3
miss ratio to represent the ratio of data prefetched directly from DRAM
by L1 over the total number of data prefetched by L1 on LFB.
This explains the outliers in SKX. SPR/EMR has the simialr issues on SQ.
Moreover, SPR/EMR does not support measuring L1/L2 prefetching offcore hit for each process.
This limitation explains the slightly worse predictions on SPR/EMR.

{\vni\bf Deployment.} To derive the DRAM slowdown model for a server and
CXL configuration, users do not need to go through the extensive
characterization process as we do because our models are robust and
independent of workloads. Thus, one could use microbenchmarks (\eg,
pointer chasing) to derive the parameters of the linear model. Below we
provide a high-level overview of the process.
There are several constants ($k_1$-$k_4$, $p$, $q$) in our prediction
model equation. They can be obtained by a set of microbenchmarks with
distinct memory access patterns.

We rely on three types of microbenchmarking workloads to derive the
model parameters. The first one is a random pointer-chasing workloads,
which imposes zero cache and store slowdowns. It can be seen as having
pure DRAM slowdown on CXL. After running it on both local and CXL
memory, we can get the overall slowdown ($S$) and calculate the DRAM
metric ($M_{DRAM}$). Due to $S_{store}$ and $S_{cache}$ being 0, $k_1$
will be $S/M_{DRAM}$.

Our next microbenchmark is a store-bound workload, \eg, one with many
\ts{malloc()}. It does not experience cache slowdown, because its
accesses do not rely on data from prefetchers. In this case,
$S=(k_1*M_{DRAM})+(k_3*M_{store})$. After the microbenchmark running on
both local and remote, $S$ is the overall slowdown. $M_{store}$ and
$M_{DRAM}$ can be obtained from the local run. Then $k_3$ could be
calculated .

To reveal cache slowdown, the third microbenchmark conducts linked-list
traversal, which requires data fetched by prefetchers. After running it
on both local and remote, $S$ can be obtained by the running time. And
$S=(k_1*M_{DRAM})+(k_3*M_{store})+(k_2*M_{cache})$. Indeed,
$(k_3*M_{store})$ should be small because it has few data allocations or
writes. $M_{DRAM}$ can be calculated with CPU counters measured on
local. Given by $S$, $k_1$, $k_3$, $M_{store}$ and $M_{DRAM}$, $k_2$ can
be obtained.
Finally, to improve accuracy, one could consider mixing up these types
of memory access and also applying linear fitting.

%% file: fig-pred.tex
\begin{figure}[t!]
\centerline{
\includegraphics[width=\columnwidth]{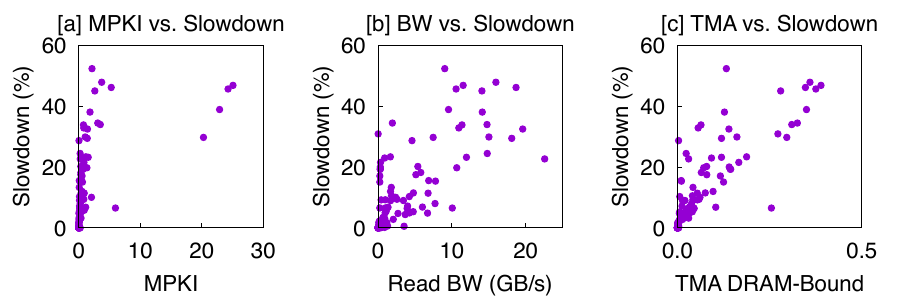}
}
\vminfive
\mycaption{fig:ncor}{Strawman prediction}{Metrics like MPKI, BW, and
TMA DRAM-Bound are not reliable CXL slowdown predictors.}
\vminten
\end{figure}

\begin{figure}[t!]
\centerline{
\includegraphics[width=\columnwidth]{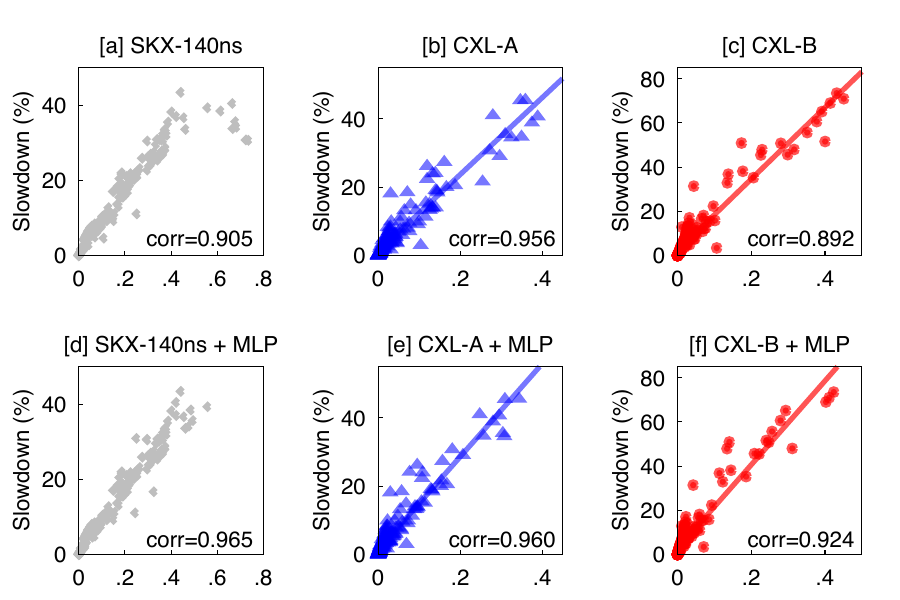}
}
\vminfive
\mycaption{fig:pm-dram}{DRAM slowdown model}{X-axis is our predictor
(discussed later) and Y-axis is measured DRAM slowdown. 182
latency-sensitive workloads are shown. (a)-(c) show the basic DRAM
model by using ``{l3-stalls/cycle}'' as the predictor for
SKX-\znuma, \cxla, and \cxlb respectively;  (d)-(f) represent the
enhanced DRAM model incorporating memory level parallelism (MLP) impact,
improving the model accuracy.}
\vminten
\end{figure}

%% file: fig-cxl-mlp.tex
\begin{floatingfigure}[r]{1.5in}
    \vspace{-0mm}
    \centering
    \hspace*{4pt}\includegraphics[width=1.5in]{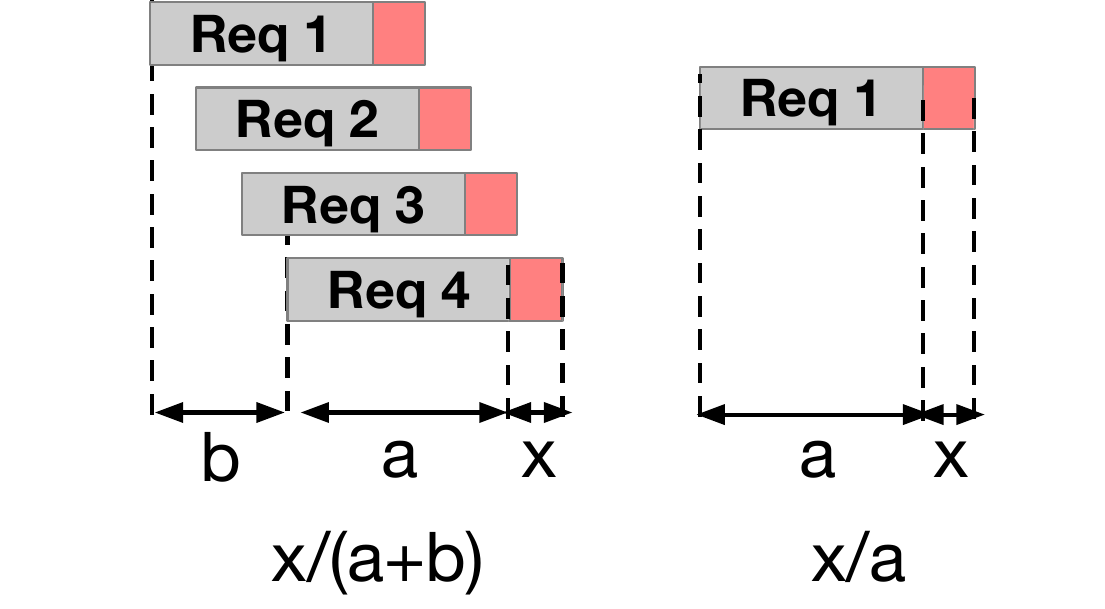}
    \vspace{-4.5mm}
    \vminfive
    \caption{\bf\small CXL MLP}
    \label{fig:mlp}
\end{floatingfigure}

%% file: tab-model-dram-err.tex

\setlength{\tabcolsep}{2pt}

\begin{table}[t]
\begin{center}
{\small
\begin{tabular}{c|ccc}
       &  \znuma   &   \cxla   &  \cxlb   \\
\hline
Pearson Correlation Coefficient       &   0.965   &   0.960   &   0.924     \\
\hline
Absolute Error within 5\%       &   92.0\%   &   94.0\%   &   78.7\%     \\
\hline
Absolute Error within 10\%      &   99.1\%   &   98.3\%   &   89.9\%    \\
\end{tabular}
}
\end{center}
\vminfifteen
\mycaption{tab:dram-err}{{DRAM slowdown prediction
accuracy}}{We can achieve 78.7\%--94\%
accuracy under 5\% misprediction target while the accuracy goes up to
89.9\%--99.1\% under 10\% misprediction.}
\end{table}

%% file: fig-pred-cache-store.tex
\begin{figure}[t!]
\centerline{
\includegraphics[width=\columnwidth]{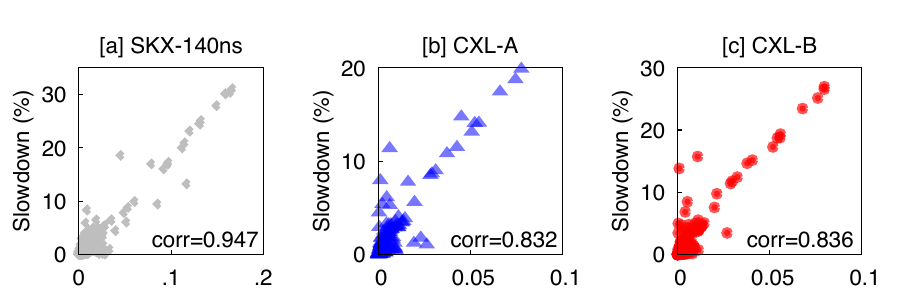}
}
\vminfive
\mycaption{fig:pm-cache}{Cache slowdown model}{X-axis is our predictor,
Y-axis is actual cache slowdown. (a)-(c) for SKX \znuma, \cxla and
\cxlb for all workloads.}
\vminten
\end{figure}

\begin{figure}[t!]
\centerline{
\includegraphics[width=\columnwidth]{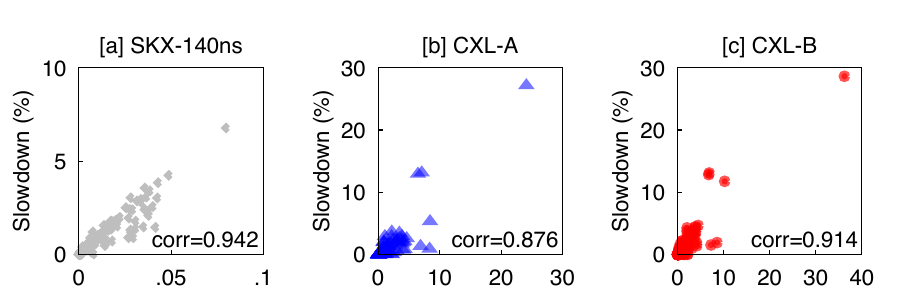}
}
\vminfive
\mycaption{fig:pm-store}{Store slowdown model}{Similar to cache slowdown model in Figure \ref{fig:pm-cache}.}
\vminten
\end{figure}

%% file: tab-model-cache-err.tex

\setlength{\tabcolsep}{2pt}

\begin{table}[t]
\begin{center}
{\small
\begin{tabular}{c|ccc}
       &  \znuma   &   \cxla   &  \cxlb   \\
\hline
Pearson Correlation Coefficient       &   0.947   &   0.832   &   0.836     \\
\hline
Absolute Error within 5\%       &   95.8\%   &   96.5\%   &   93.6\%     \\
\hline
Absolute Error within 10\%      &   99.2\%   &   98.8\%   &   98.2\%    \\
\end{tabular}
}
\end{center}
\vminfifteen
\mycaption{tab:cache-err}{Prediction accuracy of cache slowdown}{}
\end{table}

%% file: tab-model-store-err.tex

\setlength{\tabcolsep}{2pt}

\begin{table}[t]
\begin{center}
{\small
\begin{tabular}{c|ccc}
       &  \znuma   &   \cxla   &  \cxlb   \\
\hline
Pearson Correlation Coefficient       &   0.942   &   0.876   &   0.914     \\
\hline
Absolute Error within 2\%       &   97.8\%   &   93.7\%   &   95.6\%     \\
\hline
Absolute Error within 5\%      &   99.1\%   &   97.5\%   &   98.1\%    \\
\end{tabular}
}
\end{center}
\vminfifteen
\mycaption{tab:store-err}{Prediction accuracy of store slowdown}{}
\end{table}

%% file: fig-model-eq.tex
The overall slowdown model ($S$) is described below. \pone--\ptwet\ 
are the CPU counters needed for the DRAM
($M_{DRAM}$), cache ($M_{cache}$), and store ($M_{store}$) performance
predictors. $k_1$, $k_2$, $k_3$, $k_4$ are constants.

\begin{center}
{\def \hna {\hspace*{-12pt}}
{\small
\ni $S$ = $k_1 \times M_{DRAM} + k_2 \times M_{cache} + k_3 \times M_{store} + k_4$

\hna $M_{DRAM}$ = ${P_4} / {P_1} \times 1 / (p \times 1 / (P_{12}/P_{11}) + q)$


\hna $M_{cache}$ = {\small (\pthr--\pfou) / \pone \tms \psix / (\pfiv+\psix) \tms \pthit / \pfout \tms \pfift/(\pfift+\psixt)}



\hna $M_{store}$ = ${P_7}/{P_1}$
}}

\end{center}

%% file: m-inter.tex
\section{System Optimization}
\label{sec:opt}

In this section, we show \sys root-cause analysis approach and slowdown
prediction models can aid inefficiency detection and performance
optimizations under complex memory policies, such as interleaving and
tiering.

\subsection{Interleaving Characterization}

NUMA interleaving can potentially speedup bandwidth-bound workloads by
leveraging the additional CXL bandwidth alongside system memory.
Recently Linux kernel introduced weighted interleaving policy
\cite{winter.web} which allows more flexible page interleaving across
two or more NUMA nodes. Traditionally, Linux defaults to \ts{1:1} page
interleaving (\ie, \ts{MPOL\_INTERLEAVED}) where page allocations are
done in round-robin fashion across multiple NUMA nodes. Weighted
interleaving define a general \ts{M:N} interleaving ratio so that one
could use the bias to match the bandwith characteristics. For instance,
in a two-node system, weighted interleaving \ts{M:N} means that the
first \ts{M} pages are allocated from one node, followed by the next
\ts{N} pages from the other node, alternating between the two in this
pattern. However, it is not intuitive to decide the best interleaving
ratio to extract the best potential performance for certain workloads.

For bandwidth-bound workloads, it is alluring to exploit both system and
CXL bandwidth to improve performance.
Suppose system memory bandwidth is \ts{M} and CXL memory bandwidth is
\ts{N}, simply using an interleaving ratio of \ts{M:N} do not always
lead to the
best performance. A naive approach is to randomly try out \ts{M \tms\ N}
interleaving ratios which can be extremely time-consuming for
long-running workloads.
In a recent work, Caption \cite{demystifycxl.micro23} proposed
heuristics to converge over the best interleaving ratio setup but still
requires a few runs and could potentially lead to suboptimal results.

\input{fig-il-649}

We show that by adopting a performance ``slowdown'' prediction
model for interleaving similar to \sec\ref{sec:pred}, we can predict the
best page interleaving ratio for best performance, thus ``best-shot interleaving.''

{\vni\bf Offline Analysis.}
We conducted an extensive offline analysis across 100 different
local/CXL interleaving ratios (100:0, 99:1 to 0:100), for over 100 workloads. Figure \ref{fig:il649} illustrates
one such example for workload \ts{649.fotonik}. The study yields
following key insights used to develop the best-shot interleaving model.

\myfinding{
\begin{enumerate2}

    \item Non-bandwidth-bound workloads typically cannot benefit from
        (weighted) interleaving. Even for bandwidth-bound workloads, we
        observed various slowdowns (and occasional speedups) across
        different interleaving ratio settings, reflecting the combined
        impact of CXL latency and bandwidth.

    \item Various bandwidth-bound workloads have different optimal
        interleaving ratios, outperforming local DRAM performance to the
        greatest extent (there may exist a range of ratios yielding
        superior performance than local DRAM).

\end{enumerate2}
}

Our model aims to (1) predict whether a workload can
benefit from interleaving (otherwise referring to a tiering policy), (2)
predict the best interleaving ratio in one run, and (3) predict
potential performance gains.

\input{fig-il-predmodel}

\subsection{Best-Shot Interleaving Prediction}

Our slowdown breakdown method (\sec\ref{sec:rca}) can be used to analyze
and predict NUMA interleaving performance as well. For those workloads
benefiting from interleaving, the performance improvement (\ie, negative
slowdown) can still be attributed to various sources (DRAM, cache and
store).

We observe that offcore latency goes down when workload performance is
improved under efficient interleaving.
Under effective interleaving, we observe offcore latency ($L$) and
memory metric ($M$, \sec\ref{sec:pred:all}) are complementary,
where $M$ indicates the latency-impact of the memory subsystem
and offcore latency indicates memory bandwidth constraints. Thus, our
model simply adopts ($L \times M$) as the performance metric ($R$) for
interleaving performance prediction.
Similarly, it is beneficial to break down the NUMA performance
improvement or slowdown into various sources for accurate modeling. As
such, we define the following models to predict interleaving
speedup/slowdown from DRAM, cache and store, respectively.

\vspace{-2em}
\begin{align}
    R_{DRAM} &= M_{DRAM} \times L_{DRAM} \\
    R_{cache} &= M_{cache} \times L_{cache} \\
    R_{store} &= M_{store} \times L_{store}
\end{align}

We will demonstrate how $R$ can be used to predict the optimal NUMA
interleaving performance across various NUMA interleaving ratios.
Through an analysis of offcore latency when running all workloads on the
local DRAM, we identify 20 bandwidth-intensive workloads (comprising 3 SPEC workloads and
17 Llama workloads) that exhibit performance improvements with NUMA
interleaving.  We use them for our model evaluation.

{\vni\bf Evaluation on SKX-\znuma.}
Figure \ref{fig:bs-skx}a\&b show our model accuracy to derive
interleaving speedup contributions from DRAM and cache.
We also observe that for workloads with over 5\% improvement, the
optimal ratio falls within the range of 34\% to 40\% (Figure \ref{fig:bs-skx}c\&d).
Within this range, NUMA interleaving performance are roughly equally
good.
SKX2S local and \znuma bandwidth ratio is approximately $5/3$ (Table
\ref{tab:hw}). The theoretically ideal proportion of memory allocation
is $5/8$ (62.5\%) and $3/8$ (37.5\%) for local and CXL, respectively.
Consequently, the optimal ratio consistently falls within the range of
34\% to 40\% for workloads that significantly benefit from NUMA
interleaving.
For workloads with less pronounced benefits ($<$5\%), the
ratio predominantly ranges between 20\% and 36\%.
The store model ($R_{store}$) is trivial for most workloads, thus, we
omit store interleaving slowdown analysis here.

{\vni\bf Evaluation on \cxla.}
Figure \ref{fig:bs-cxla} demonstrates the linear relationship between
the best interleaving ratio and our predictor ($R$).
Compared to Caption \cite{demystifycxl.micro23}, our approach
greatly simplify the process in an automatic way.  Overall, best-shot
interleaving achieves 1-13\% performance improvement compared to
local-DRAM on 20 workloads (upper bound is limited by the relatively low
CXL bandwidth).

{\vni\bf Latency-bound workloads.} For workloads not constrained by bandwidth,
the performance varies approximately linearly across different ratios.
Although we could not gain interleaving benefits, interestingly, our
model still allows us to predict the slowdown under a givn interleaving
ratio $x$ using a simple mode such as $x \times \sum_{DRAM, cache,
store}S_i$, where $x$ represents $N/(N+M)$ (with $N$ denoting the remote
node ratio and $M$ representing the local node ratio), and $S_i$ denotes
the slowdown on CXL for different hardware components.

\myimplict{Weighted page interleaving can be used to improve performance
for certain bandwidth-bound workloads under local and CXL memory.
However, the optimal interleaving ratio varies across different
workloads and the degree of performance improvements also differs.
Best-shot interleaving can help predict the best interleaving ratios to
achieve optimal performance and predict the precise amount of performance
gains.}

\myrecommend{For bandwidth-bound workloads, the users can rely on our
best-shot interleaving policy to run their workloads using the best
setup for optimal performance.}

%% file: fig-il-649.tex
\begin{floatingfigure}[r]{1.5in}
\vminfive
\centerline{
\includegraphics[width=1.4in]{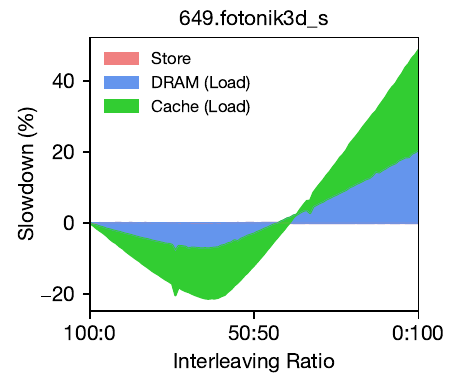}
}
\vminfive
\mycaption{fig:il649}{Weighted interleaving performance}{}
%
%
\end{floatingfigure}

%% file: fig-il-predmodel.tex
\begin{figure}[t!]
\centerline{
\includegraphics[width=1.0\columnwidth]{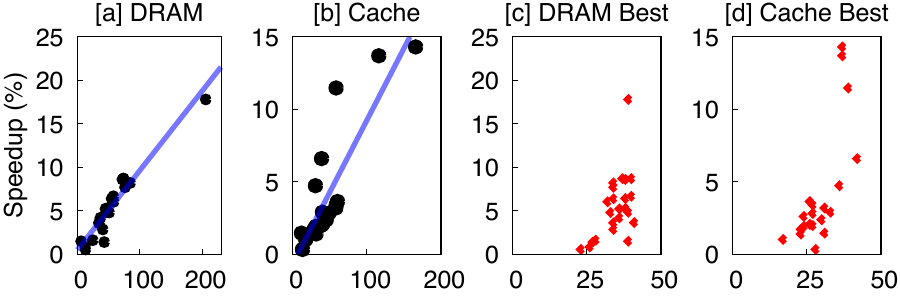}
}
\vminfive
\mycaption{fig:bs-skx}{Best-shot interleaving for SKX}{In (a) and (b),
    X-axis is our predictor ($R$), and Y-axis is the (predicted) speedup
    from DRAM and cache respectively. The black circles are offline
    optimal interleaving results. In (c) and (d), X-axis is the best
interleaving ratio assigned to \znuma and Y is the actual
interleaving performance speedup sourced from DRAM and cache. The best
interleaving ratio on SKX ranges from 34\%-40\%.}
\vminfive
\end{figure}

\begin{figure}[t!]
\centerline{
\includegraphics[width=1.0\columnwidth]{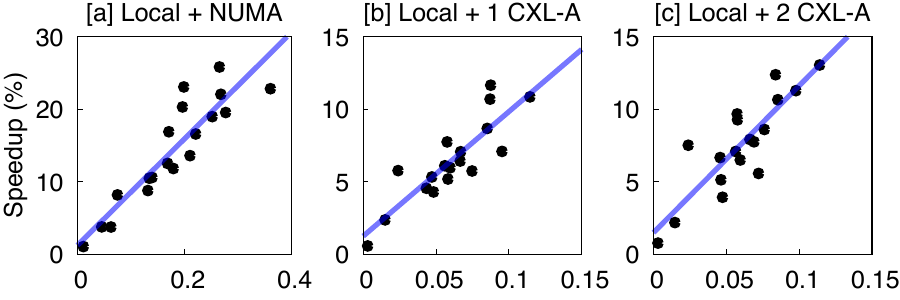}
}
\vminfive
\mycaption{fig:bs-cxla}{Best-shot interleaving model for \znuma, 1 and
2 \cxla device(s)}{(a)-(c) show that best-shot interleaving model is
accurate for both \znuma and real CXL devices. Under 20 workloads in
\znuma, 1 and 2 \cxla devices, best-shot interleaving can accurately
predict and achieve performance gains of 2-21\%, 1-13\%, 1-26\%,
respectively.}
\vminten
\end{figure}

%% file: m-tier.tex
\subsection{Tiering Characterization}
\label{sec:opt:tier-char}

We now show \sys root-cause breakdown analysis and performance models
can be applied to tiering systems to dissect inefficiencies in tiering
systems.

Existing tiering designs implicitly treat each LLC miss equally in terms
of their contribution to system performance and heavily rely on LLC
misses as the primary technique for sampling hot pages as migration
candidates.
However, our slowdown breakdown analysis (\sec\ref{sec:rca}) has
demonstrated that LLC misses (or their rate, \ie, bandwidth) cannot
reliably serve as a performance predictor/metric. This is because LLC
misses caused by prefetching or RFO may not directly impact system
performance. For instance, a prefetched cacheline may end up not being
used. Instead, we assert that the rate of LLC stalled cycles and other
stall cycle-related events are more accurate measures to gauge
and predict system pressure.

Nonetheless, current tiering policies overlook this nuance and
indiscriminately assume that high rates of LLC misses (or equivalently,
DRAM traffic) inevitably result in performance degradation,
inadvertently promoting excessive pages to local DRAM. This approach
carries two downsides. First, migrating a large number of pages incurs
non-negligible overheads, further compromising workload performance.
Second, the assumption that these pages merit promotion to the fast tier
(local DRAM) is unfounded, as they may not induce significant slowdowns.
In combination, these factors lead to suboptimal tiering performance.

\input{fig-tiering-example}

{\vni\bf Characterizing tiering inefficiencies.}
We now use the prior analysis to reason about potential inefficiencies
in tiering systems with a realistic workload, namely tc-twitter.
In Figure \ref{fig:tiering-example-tc-twitter}a, we applied our
``slowdown'' prediction models to analyze tc-twitter slowdowns-over-time
under CXL.
Here, we apply our model to a period of the workoad executions (\eg, every 1B
instruction interval). Similar to
previous workload level DRAM-contributed slowdown prediction, applying
the LLC-stalls together with MLP factor delivers better prediction, even
at very fine-granularity (pink line is very close to blue line).
We can see that the most significant DRAM-introduced slowdowns for
tc-twitter occur during the final phase of the execution (3rd-4th
billion instruction period).
Upon further profiling of the LLC miss rate of tc-twitter over time (Figure
\ref{fig:tiering-example-tc-twitter}b), we found that
the bulk of LLC misses occur during the initial phase.
That said, the substantial number of LLC misses during the initial phase
do not contribute significantly to the workload performance as the rest
of phases.

However, existing tiering designs operate under the assumption that
performance degradation correlates positively with memory access rates.
Consequently, they tend to aggressively ``detect/scan'' and migrate ``hot''
pages (both promotion and demotion).

\myfinding{As a result, it causes two potential problems: (1) ``hot'' pages are
wrongly detected, \ie, the migration of these seemingly hot pages does
not lead to an overall performance enhancement as they don't cause CPU
stalls by default; (2) As a result of the wrong hot page detection, it
triggers unnecessarily high number of page migrations, which inversely
degrade system performance (page-level migrations are long-latency and
blocking operation in nature, causing high overhead). Combined, these
would render tiering systems underperform compared to no-tiering.}

Using TPP \cite{tpp.asplos23} as an illustrative example, we demonstrate
how such memory tiering designs can exacerbate overhead and result in
wrong page promotion decisions. In Figure
\ref{fig:tiering-example-tc-twitter}d, the blue line shows the page
promotion rate over time, which shows similar patterns as the LLC misses
over time in Figure \ref{fig:tiering-example-tc-twitter}b.
Correspondingly, a peak of 50,000 pages/s were observed around time 30s.

\myfinding{We define a new metric called ``amortized offcore latency''
considering both memory latency and MLP impact to capture the impact
of CXL memory accesses to workload performance (details ommitted).
And we find it to be able to capture workload performance very
well.}

In Figure \ref{fig:tiering-example-tc-twitter}c, we show that the
``amortized offcore latency'' during the initial phase remains notably
low, indicating significant read request overlappings during the period.
This overlapping mitigates performance degradation even in the presence
of high LLC miss stalls, as many memory accesses,
despite being affected by increased remote memory latency, are concealed
by other parallel reads.

Further validation in Figure \ref{fig:tiering-example-tc-twitter}a and
Figure \ref{fig:tiering-example-tc-twitter}b confirms that
the high LLC misses during the initial phase result in marginal DRAM
slowdown that is not as pronounced as observed during the final phase of
the workload.

\subsection{\alto: Adaptive Layered Tiering Orchestration}
\label{sec:opt:alto}

Our optimization is straightforward: limiting page promotions when the
overlapping effect of memory accesses is evident.
To this end, we propose {\bf \alto, an adaptive layered tiering
orchestration} scheme, built on top of TPP, to demonstrate the efficacy
of our method.
We chose TPP as it is the latest tiering effort tailored for CXL while
alternatives like Hemem \cite{hemem.sosp21} and Memtis
\cite{memtis.sosp23} primarily target persistent memory.
Additionally, it's worth noting that page sampling (\eg, Intel PEBS), an
enabling technique for Hemem and Memtis does not support CXL yet.

We implement \alto by constraining the page promotion rate
proportionally to the ``amortized offcore latency'' based on two
thresholds.
Specifically, if the ``amortized offcore latency'' (\sec\ref{sec:pred:dram}) falls below a lower
bound, \eg, 40 cycles, we disable
page promotion to account for the evident memory access overlapping
effect. Otherwise, if it exceeds the upper threshold, \eg, 100 cycles,
we do not limit page promotions. Both the lower bound and upper bound
thresholds can be derived offline using a
microbenchmark similar to \sec\ref{sec:pred:all}.

In between, we gradually reduce page promotion rate as amortized offcore
latency decreases, using a default 5-step interval.
In our implementation, we achieve this by periodically ignoring
potential promotion page candidates within small sets of pages. For
instance, if we aim to allow 20\% of TPP-identified candidate pages to
be promoted, we allow the first two pages of every 10 pages to go
through.

To monitor the ``amortized offcore latency'', we collect PMU counters
periodically, \eg, every 1s.  Subsequently, we calculate the amortized
offcore latency based on these counters, enabling us to dynamically
adjust the page promotion rate based on the observed latency. Our
user-level tool is lightweight and imposes no additional overheads. The
kernel side only involves \roughly 30 LOC changes to Linux MM migration
policies. Reading a couple of PMU counters is extremely lightweight.
Alto reads only 5 PMU counters every second, imposing almost zero
overhead.

\input{fig-tiering-tpp}

{\vni\bf \alto Evaluation.}
We test \alto with 8 workloads, including graphs, ML and SPEC, comparing
it with TPP and three additional settings: workloads backed by all local
memory (Local), CXL memory (CXL), and default Linux hybrid local/CXL
memory without tiering (default Linux). Since TPP performance is
sensitive to the fast-tier memory size, we configure the local memory
size to be large enough to accommodate the entire workload working set
(profiled offline).

Workload working set (WSS) means the part of memory footprint which is
actively accessed during workload runtime.  We estimated WSS using
heatmaps obtained via offline PEBS-based LLC-miss sampling (high
sampling rate at 100 for accuracy).  For each workload, we set its local
DRAM to be slightly larger than its working set size (WSS), and CXL is
used for the remaining memory footprint (\ie, total memory footprint
minus WSS). CXL memory is constantly accessed by the workloads as
first-touch doesn't guarantee all the hot pages (in WSS) are initially
placed in local memory, Figure \ref{fig:tiering-example-tc-twitter}d
showcases heavy page promotions from CXL to local memory for tc-twitter.
The gap between Linux and Local in Figure
\ref{fig:tiering-example-tc-twitter} stems from the accesses to CXL.  We
argue our local/CXL setup is fair to evaluate TPP as TPP performs much
worse when more (slow) CXL memory is used, under which case Alto can
actually improve TPP up to 9\tms\ (not shown).

In our evaluation, TPP typically underperforms default Linux due to
erroneous page migration decisions and the resulting excessive overhead.
We present all the results in Figure \ref{fig:alto}. \alto demonstrates
improved performance compared to default Linux for workloads such as
bc-twitter (+16\%), bc-urand (+18\%), and tc-kron (+3\%). This
enhancement stems from the fact that memory tiering can achieve better
performance when it migrates correct pages. \alto enables TPP to
constrain unnecessary page migrations by using an accurate
performance metric, thereby aligning its behavior more closely with
optimal performance scenario.

In detail, \alto demonstrates a performance improvement over TPP ranging
from 0.7\% to 177.5\%. The most notable enhancement is observed in
workload GPT-2, attributed to its highly parallel memory accesses and
the substantial migration overheads in TPP. For bc-twitter, TPP even
exhibits a 62\% slower performance compared to CXL, while \alto
significantly enhances TPP's performance.
Workload tc-kron experiences the least performance improvement under
\alto, primarily because only a small portion of it exhibits overlapped
memory accesses.
\alto outperforms Linux for 3 out of the 8 workloads in Figure
\ref{fig:alto} by 3\%, 11\%, and 14\% while only slightly
underperforming by 3-6\% for the rest. Note that, in most cases, tiering
designs such as TPP/AutoNUMA lose to first-touch/Linux as tiering
becomes more sensitive to page migration overhead given the small
latency gap (1.9-2.4\tms) between CXL/local memory.

It is an unfortunate (and maybe surprising) fact that
first-touch/Linux under CXL is actually better than many (if not all)
state-of-the-art tiering policies.  According to our evaluations, TPP,
AutoNUMA, and Nomad \cite{nomad.osdi24} loses to Linux by up to 181\%, 22\%, and 50\%,
respectively. Nomad authors also acknowledged in their paper (Section
4.2) that No-Migration (aka, Linux) performance exceeds (all) tiering
solutions. This is because CXL latency is only 1.9-2.4\tms\ that of
local-DRAM (for CXL-A,B,D) and the overhead of page migration can easily
outweigh its benefits if migration policy is not carefully designed.

\myimplict{More broadly, we think tierability needs to be revisited in
the CXL era. \alto's advantage over Linux/First-touch (even just)
for some workloads calls for the need for principled approaches like
ours to {\bf (1)} diagnose and characterize tiering inefficiencies
beyond hot/cold separation, and {\bf (2)} revisit tiering policies
designs to reduce migration overheads and focus on migrating
performance-sensitive pages.}

We utilize \alto to demonstrate how a performance metric
from \sys insights can significantly aid in identifying inefficiencies
and enhancing existing tiering system performance with minimal changes.
While \alto does not directly address the challenge of accurately
sampling the most performance-critical hot pages for migrations,
orchestrating the page migration rates indeed helps mitigate the overhead
of incorrect migrations across a range of workloads. Additionally, we
believe that \sys's CPU-stall-based approach could further
improve hot page sampling accuracy.

%% file: fig-tiering-example.tex
\begin{figure}[t!]
    \begin{subfigure}[t]{0.495\linewidth}
        	\centering
        	\includegraphics[width =\textwidth]{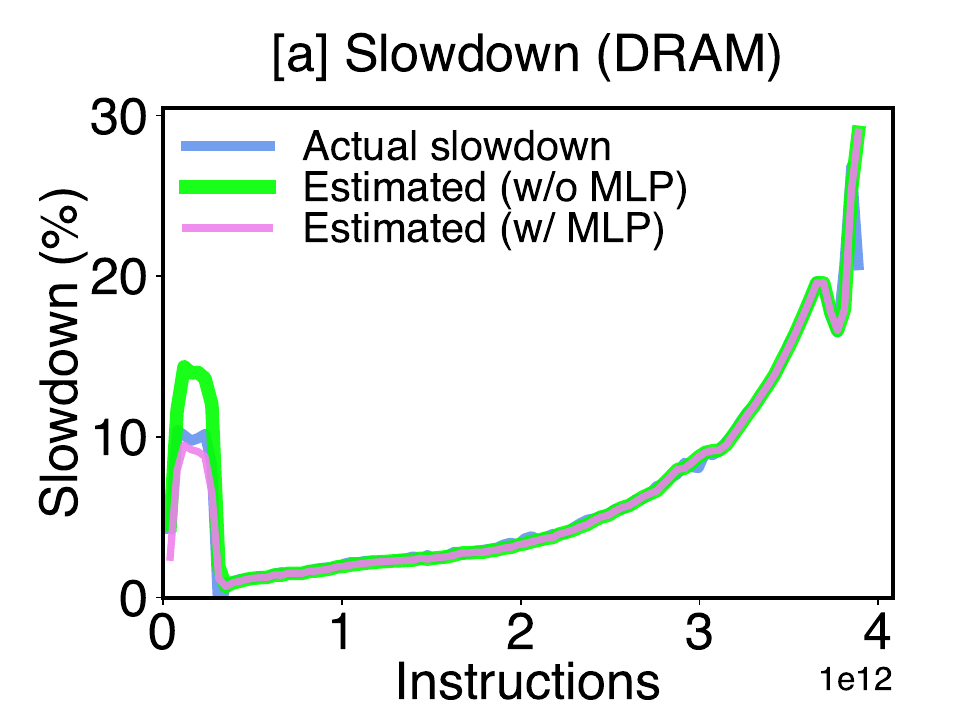}
            \vspace{-5mm}
            \label{fig:tiering-example-tc-twitter-est}
    \end{subfigure}
    \hfill
    \begin{subfigure}[t]{0.495\linewidth}
        	\centering
        	\includegraphics[width =\textwidth]{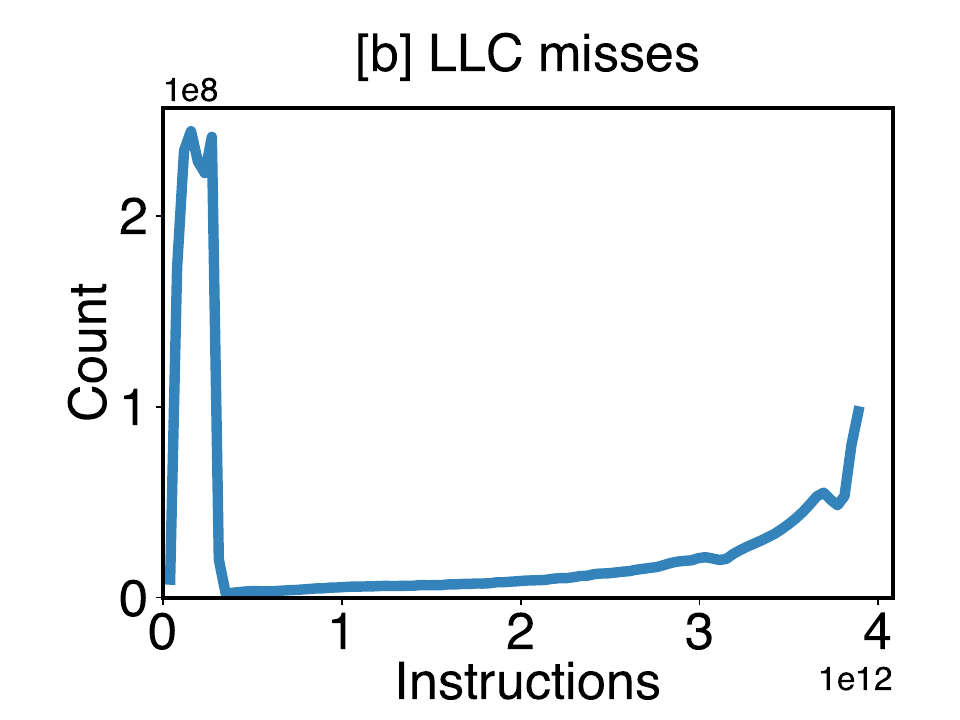}
            \vspace{-5mm}
            \label{fig:tiering-example-tc-twitter-llc}
    \end{subfigure}
    \hfill
    \begin{subfigure}[t]{0.495\linewidth}
        	\centering
        	\includegraphics[width =\textwidth]{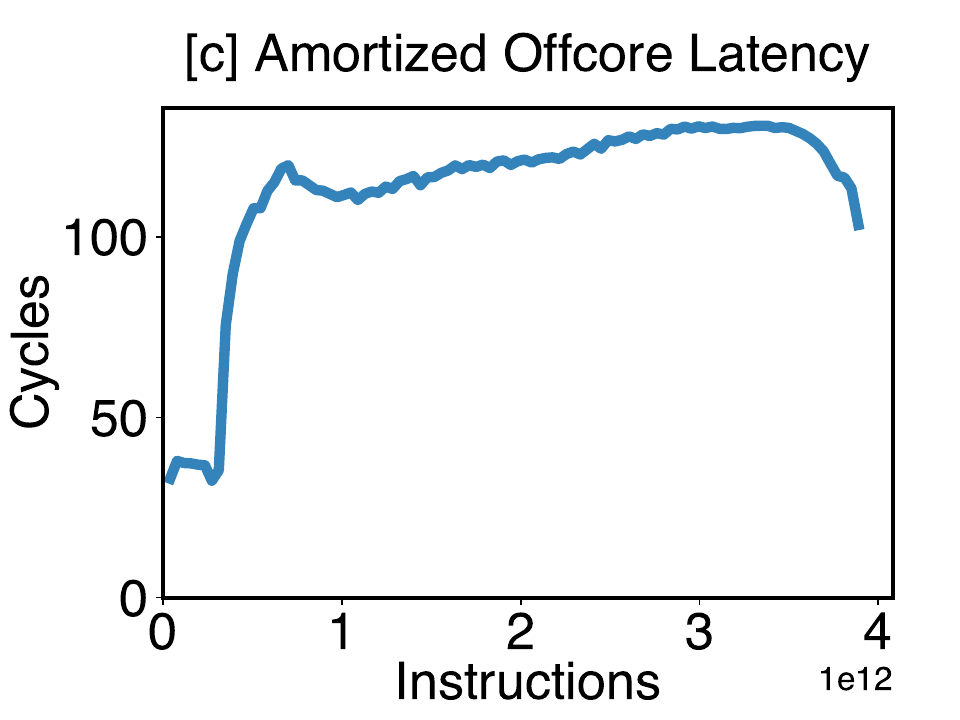}
            \vspace{-5mm}
            \label{fig:tiering-example-tc-twitter-lat}
    \end{subfigure}
    \hfill
    \begin{subfigure}[t]{0.495\linewidth}
        	\centering
        	\includegraphics[width =\textwidth]{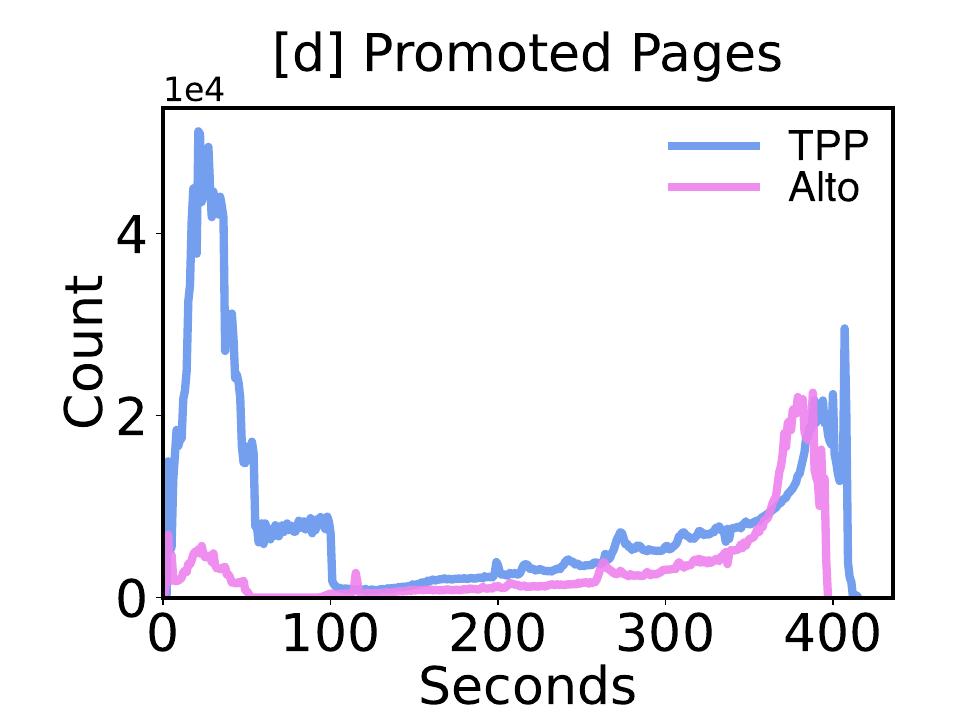}
            \vspace{-5mm}
            \label{fig:tiering-example-tc-twitter-pro}
    \end{subfigure}
    \vminfifteen
    \mycaption{fig:tiering-example-tc-twitter}{Tiering performance
    characterization}{LLC miss is not a good predictor for guiding
        memory tiering decisions. Our \ts{L3-stalls+MLP} metric is more
    accurate. Note that over 99\% of tc-twitter's slowdown
    originates from DRAM.}
    \vminten
\end{figure}

%% file: fig-tiering-tpp.tex
\begin{figure}[t!]
\vspace{-0mm}
\centering
\includegraphics[width=\linewidth]{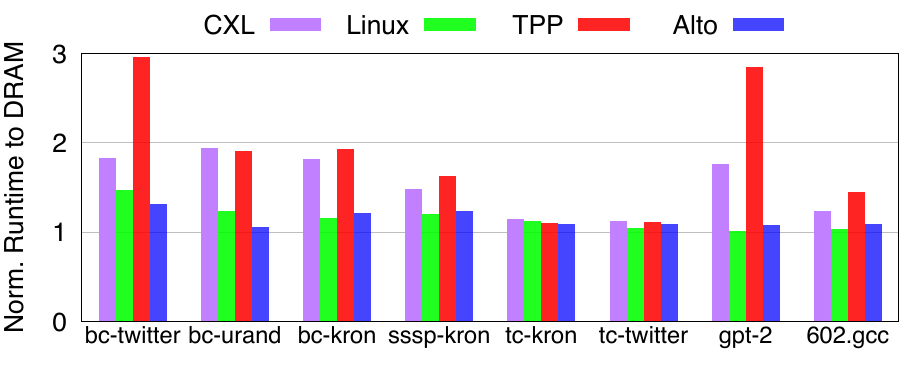}
\vminten
\vminten
\mycaption{fig:alto}{\alto performance vs. TPP}{X-axis is 8 different
workloads under test, Y-axis is normalized workload runtime to Local
DRAM. \alto can outperform TPP by 0.7-177\%.}
\vminten
\end{figure}

%% file: dis.tex
\section{Discussion}
\label{sec:dis}

{\ni\bf \sys implications.} While the study primarily focuses on CXL
devices, the high prediction accuracy on \znuma indicates a pathway to
performance observability, explainability, and predictability of general
memory systems, relying solely on simple combinations of lightweight
performance counters. Stemming from an offline performance breakdown
analysis, \sys performance models turn out to be workload-independent,
accurate, robust, lightweight, simple, universal, and explainable.  Our
models are validated across \numCxlDevs\ different CXL devices and
\numCpus\ processor platforms, demonstrating the broad applicability of
our model and the effectiveness of our modeling methodology. This paves
the way for potential generalization. The simplicity of \sys models
should facilitate both offline and online usage. Our performance models
can potentially serve as general performance metrics/predictors for
various tasks, such as workload/VM resource management and task
scheduling.
\sys identifies key performance metrics that we envision can guide
numerous system task optimizations, including hybrid memory policies
integrating the benefits of interleaving and tiering, as well as new
tiering policy designs such as improved hot/important page sampling.

{\vni\bf CXL performance predictability.} Our prediction models'
deterioration from \znuma to \cxla or \cxlb indicates that
\cxlb's worse tail latency also corresponds to the reduced
predictability of our corresponding performance prediction models
compared to \znuma and \cxla. This trend may worsen when future
CXL-attached persistent memory or NAND Flash devices emerge.  Addressing
this challenge requires collaborative efforts from CPU, CXL device
vendors, and OS/software developers to build QoS-aware and tail-tolerant
software and hardware memory systems.

Additionally, CXL tail latencies also adversely affect academic CXL
research based on emulation/simulation, such as \znuma, given the
current scarcity of CXL devices. Properly modeling and simulating CXL's
intricate performance characteristics are essential to ensure a true
reflection of real hardware characteristics.

{\vni\bf Workload co-location:} We validate that our models work for
colocated applications as well (\eg, multiple instances of various CPU
2017 workloads).

{\vni\bf Future-Proofing.} Future CXL devices will significantly
improve bandwidth and somewhat improve latency. We anticipate our major
indings and optimizations to remain valid.
\begin{enumerate2}

\item CXL tail latencies are likely to persist due to various
    performance-functionality trade-offs in CXL controller
    implementations/optimizations, such as request scheduling, thermal
    management, QoS, and Reliability, Availability, and Serviceability
    (RAS). For instance, PCIe 6 will require thermal throttling,
    which could potentially worsen tail
    latencies \cite{pcie6tt.web, pcie6hot.web}. Additionally, with future CXL devices connected through
    CXL switches, the additional hops and potentially slower media
    (PM/Flash) will further increase the chances of latency
    unpredictability.

\item Future CXL workload slowdowns will likely be smaller than those in
    Figure \ref{fig:wcdf}a. Increased CXL bandwidth will benefit
    bandwidth-bound workloads, alleviating the 2-6\tms\ slowdowns seen
    in Figure \ref{fig:wcdf}a due to low-bandwidth per CXL device in our
    setup. Further latency reductions will improve the performance of
    latency-sensitive workloads, such as cloud
    applications, approaching NUMA
    performance. This is already evident with CXL-D\textsuperscript{*}
    (hardware-interleaving across two CXL-Ds, \lmt 100GB/s bandwidth,
    green line) in Figure \ref{fig:wcdf}a, where bandwidth is no longer
    a bottleneck, similar to NUMA (black line). However, the latency gap
    between CXL and local memory persists. Mitigating slowdowns from CXL
    latencies will remain challenging without software/hardware
    optimizations, underscoring the need for detailed studies to
    characterize, analyze, model, and optimize performance to match
    local DRAM.

\item Our performance modeling approach will remain valid with improved
    CXL performance, and we expect our CXL prediction models
    (\sec\ref{sec:pred}) to become more accurate, approaching the
    accuracy of zNUMA.

\item Our best-shot interleaving policy can further benefit
    bandwidth-intensive workloads such as HPC applications, by enabling
    them to exploit the higher aggregate system memory bandwidth.

\item We expect \alto to be more effective compared to state-of-the-art
    tiering policies, as their migration overhead will become more
    apparent when the latency gap between CXL and local memory narrows.
    For instance, our \alto evaluations on zNUMA (ideal-CXL) show an
    improvement to TPP up to 248\% (not shown in the paper),
    significantly higher than the 177\% improvement for \alto on current
    real CXL.

\end{enumerate2}

%% file: rel.tex
\section{Related Work}
\label{sec:rel}

{\ni\bf CXL-based memory disaggregation.} Memory disaggregation
\cite{nimblepage.asplos19, tpp.asplos23, hemem.sosp21, memtis.sosp23,
tmo.asplos22, pond.asplos23, cxlpool.ieeemicro23, lmpool.hotnets23} is a
promising technique to improve memory resource utilization, which
recently becomes more practical thanks to CXL's cache coherent
interface. CXL-based systems \cite{cxl-shm.sosp23, pond.asplos23} need
to address various aspects of memory management, including performance
predictability. Our large-scale study contributes to a deep
understanding of CXL performance implications, potentially motivating
tailored management schemes to align with CXL performance
characteristics for its imminent deployment.

{\vni\bf Memory characterization.} While DRAM
characteristics have been extensively studied and modeled
\cite{cpistud.iiswc20, memanalysis.memsys20, memlatstd.iiswc15,
dramvar.sigmetrics16, demystifydram.sigmetrics19}, the introduction of
CXL prompts a reevaluation due to its unique performance
characteristics. For instance, we unveiled CXL tail latency in the range
of 100s of nanoseconds which is much larger than DRAM chip-level latency
variations.
Caption \cite{demystifycxl.micro23} is one of the first works
characterizing real CXL devices, revealing measurement
results of microbenchmarks and Redis/DLRM-like workloads. Due to the
black box nature of CXL devices, Caption's analysis of workload
performance is heavily reliant on speculations. While facing similar
challenges, we purposely focused on different goals in our work: a much
larger set of offline workload characterizations to reveal the detailed
CXL impact on CPU pipelines, validated to be accurate, which further
enabled us to develop an accurate performance prediction model.
This model can be used for CXL memory management optimizations in
interleaving and tiering scenarios. Our finding on CXL tail latencies,
to the best of our knowledge, is a first in the community, and we
carefully designed experiments to quantify its impact. Caption also
contributes an algorithm to derive a good interleaving ratio for
bandwidth-bound workloads; however, Caption relies on a heuristic
approach that requires running the workload multiple times (\eg, 4--10
repeated runs) to converge on the result by relying on empirical
metrics (\eg, L1 miss latency). Our best-shot interleaving policy is
inspired by Caption design and shares similar goals. However, we achieve
more ambitious goals to predict both the optimal performance and
weighted interleaving ratio in one run, guided by a systematic reasoning
which is more accurate.

{\vni\bf Memory tiering.} Memory tiering \cite{autotiering.atc21,
autonuma.web, nomad.osdi24, tpp.asplos23, hemem.sosp21, memtis.sosp23}
typically relies on page table scanning, NUMA page-fault hints, and
hardware event sampling (\eg, Intel PEBS) to detect hot/cold pages,
treating all memory accesses to DRAM equally without considering their
relative contribution to workload performance in terms of CPU stalls.
Although our work is not a typical tiering paper, our prediction models
are shown to be useful in understanding inefficiencies in tiering and
enhancing its performance. We hope that our findings and insights will
guide the development of next-generation tiering policies, as we have
demonstrated using the case of \alto in \sec\ref{sec:opt:alto}.

{\vni\bf Performance prediction:} Effective performance predictors,
whether based on heuristics or machine learning, are crucial for system
resource management and scheduling decisions.
TMO \cite{tmo.asplos22} utilizes the PSI metric to guide tiering choices
across multiple types of memory backends, measuring the amount of lost
work due to resource shortages.
Pond \cite{pond.asplos23} employs an ML-based latency-sensitivity
predictor to guide pool memory allocations.
Caption \cite{demystifycxl.micro23} combines three metrics: L1 miss
latency, DRAM latency, and IPC, to converge on the best NUMA
interleaving ratio progressively.
Our work shares similar aspirations but aims to identify a fundamental
performance metric that is thoroughly reasoned and validated to be
accurate. The novel combinations of a few performance counters in \sys
make it simple and lightweight. We believe our work is complementary to
parallel explorations of new performance prediction methods with many
potential use cases.
For example, \sys models could potentially serve as a simple and
accurate replacement, \eg, for Pond's \cite{pond.asplos23} ML models,
due to their simplicity and high accuracy.

%% file: conclude.tex
\section{Conclusion}
\label{sec:conc}

In this paper, we present \sys, the largest-scale CXL memory performance
characterization conducted on a combination of hundreds of real-world
applications and multiple hardware CXL and memory configurations. Our
study unveils new findings regarding CXL performance characteristics,
contributing novel insights to the community. Importantly, the
characterization results enable a root-cause analysis for sub-\us memory
latencies, leading to our most significant contribution: memory system
performance prediction models built on just over ten performance
counters. We demonstrate that our approach to derive the model and the
model itself are useful in real-world interleaving and tiering
scenarios. We plan to open-source \sys and hope to inspire more research
in this direction to better understand and manage CXL implications for
efficient system designs.

%% file: ack.tex
\section{Acknowledgments}

%

We thank Yuyue Wang, Hansen Idden, and Shoaib A. Qazi for their
assistance in setting up the experimental environment and conducting
some of the initial experiments. This work was supported by funding from
the NSF (grant numbers
CNS-2339901
and CNS-2312785),
as well as gift and contract funding from Samsung.
